\documentclass[nohyper,12pt,letterpaper]{JHEP3}
\usepackage{graphicx,epsfig,youngtab}

\author{Vincent De Comarmond$^1$, Robert de Mello Koch$^{1,2}$ and Katherine Jefferies$^{1}$\\
$^{1}$ National Institute for Theoretical Physics,\\
Department of Physics and Centre for Theoretical Physics,\\ 
University of the Witwatersrand,\\ 
Wits, 2050, South Africa\\
\qquad\\
$^{2}$Stellenbosch Institute for Advanced Studies,\\
Stellenbosch, South Africa\\
\qquad\\
E-mail: \email{Vincent.Decomarmond@students.wits.ac.za, robert@neo.phys.wits.ac.za, Katherine.Jefferies@students.wits.ac.za}}

\abstract{
The large $N$ limit of the anomalous dimensions of operators in ${\cal N}=4$ super Yang-Mills theory 
described by restricted Schur polynomials, are studied.
We focus on operators labeled by Young diagrams that have two columns (both long) so that the 
classical dimension of these operators is $O(N)$. 
At large $N$ these two column operators mix with each other but are decoupled from operators with $n\ne 2$ columns.
The planar approximation does not capture the large $N$ dynamics. 
For operators built with 2, 3 or 4 impurities the dilatation operator is explicitly evaluated. 
In all three cases, in a certain limit, the dilatation operator is a lattice version of a second derivative, with the lattice 
emerging from the Young diagram itself.
The one loop dilatation operator is diagonalized numerically. 
All eigenvalues are an integer multiple of $8g_{YM}^2$ and there are interesting degeneracies in the spectrum. 
The spectrum we obtain for the one loop anomalous dimension operator is reproduced by a collection of harmonic oscillators. 
This equivalence to harmonic oscillators generalizes giant graviton results known for the BPS sector and further implies that
the Hamiltonian defined by the one loop large $N$ dilatation operator is integrable.
This is an example of an integrable dilatation operator, obtained by summing both planar and non-planar diagrams.}

\preprint{WITS-CTP-061}

\title{Surprisingly Simple Spectra}

\keywords{Giant Gravitons, AdS/CFT correspondence, super Yang-Mills theory}

\def \Tr{\mbox{Tr\,}}

\begin{document}

\section{Introduction}

In the last few years interesting new progress has been made in the study of the dynamics of multimatrix models.
Starting from the remarkable observation\cite{Corley:2001zk} that the Schur polynomials are a complete basis of 
gauge invariant operators which diagonalize the two point function of the free ($g_{YM}^2=0$) super Yang-Mills theory,
similar bases have been found for multimatrix 
models\cite{Balasubramanian:2004nb,de Mello Koch:2007uu,de Mello Koch:2007uv,Kimura:2007wy,Bekker:2007ea,Brown:2007xh,Bhattacharyya:2008rb,Brown:2008rr,Kimura:2008wy,Kimura:2009wy,Ramgoolam:2008yr}.
For these bases, the two point function is diagonal and known exactly as a function of $N$ (but of course, at $g_{YM}^2=0$).
The fact that the $N$ dependence is known exactly suggests that these results will be useful for going beyond the planar
approximation.

When would this be needed? Often ``the planar limit'' and ``the large $N$ limit'' are taken as synonyms. This is not, in
general, accurate. For example, if we imagine computing the two point correlator of an operator with a bare dimension $\Delta$
of most $\Delta\sim J$ with ${J^2\over N}\ll 1$, then summing the planar diagrams will capture the large $N$ limit.
For operators with a dimension larger than this, exploding combinatoric factors overpower the nonplanar (${1\over N^2}$) suppression
and the planar approximation is completely ineffective\cite{Balasubramanian:2001nh}. 
In this scenario, to get the correct large $N$ limit, it is necessary to sum a lot more than just the planar diagrams. 
On general grounds we expect the large $N$ limit to be simpler than the full theory\cite{'tHooft:1973jz}. 
The planar diagrams are a small subset of all possible diagrams, so that it is quite natural to expect that summing 
only the planar diagrams will give a much simpler problem. Why should the large $N$ limit be simple when one needs to sum much more
than just the planar diagrams? The answer to this question will probably not be general, but rather will depend on the specific
dynamical problem considered and must be answered case by case. A very pedestrian approach is simply to compute the large $N$ limit
and then to look for simplifications. This has been accomplished\cite{Koch:2008ah,Koch:2008cm,Koch:2009jc} in a number of interesting 
examples including LLM geometries \cite{Lin:2004nb,Balasubramanian:2005mg} and the near horizon geometry of a bound state of giant 
gravitons\cite{shahin}. {\sl The results are remarkably simple}. Indeed, as an example, for ${1\over 2}$ 
BPS-correlators in the presence of $M$ giant gravitons with $M$ of order $N$, \cite{Koch:2008ah,Koch:2008cm,Koch:2009jc} showed 
that the usual ${1\over N}$ expansion is replaced by a ${1\over M+N}$ expansion. Further, if one expands the exact correlators
(which because they are ${1\over 2}$ BPS do not depend on $g_{YM}^2$ but only on $N$ or $N+M$) the expansion coefficients for correlators
in the background of $M$ giants are exactly the same as the expansion coefficients for correlators with no giants present! This
remarkably simple result was confirmed holographically\cite{Maldacena:1997re} by matching to graviton dynamics in the LLM geometries
using the formalism of \cite{Skenderis:2007yb}. For near-BPS operators corresponding to BMN loops\cite{Berenstein:2002jq} 
it was argued in \cite{Chen:2007gh,us,hai1}
that the usual 't Hooft coupling $g_{YM}^2N$ is replaced by the effective 't Hooft coupling $g_{YM}^2 (N+M)$. For additional
interesting related studies see\cite{shahin,hai2}.

In this article we will consider the problem of computing the anomalous dimension of an operator with a bare dimension of order $N$.
To answer this question, we need to go well beyond the planar limit; we find the methods and approach 
of\cite{Balasubramanian:2004nb,de Mello Koch:2007uu,de Mello Koch:2007uv,Kimura:2007wy,Bekker:2007ea,Brown:2007xh,Bhattacharyya:2008rb,Brown:2008rr,Kimura:2008wy,Kimura:2009wy}
once again surprisingly effective. The results again exhibit a remarkable simplicity - the spectrum of anomalous dimension can be matched to the
spectrum of a set of oscillators! We see once again that the large $N$ limit is indeed a simple limit.

The operators we consider, restricted Schur polynomials, will be built using $O(N)$ $Z$s 
and 3 or 4 ``impurities'' ($Y$s), where $Z$ and $Y$ are complex adjoint 
scalars of ${\cal N}=4$ super Yang Mills theory. The case of operators with two impurities was studied in\cite{Koch:2010gp}. The dilatation operator
when acting on a restricted Schur polynomial, produces terms that have a combination $ZY-YZ$ appearing. In \cite{Koch:2010gp}, the techniques of
\cite{de Mello Koch:2007uv,Bekker:2007ea} were used to separate the $Z$ and the $Y$ and then write the results as a linear combination
of restricted Schur polynomials. This method is very cumbersome as it involves the inversion of a matrix. This must be done analytically
so its tedious (for the case of two impurities one must invert a $6\times 6$ matrix). More than two impurities was effectively out of
reach. In this article we develop a new formula (in section 2) which avoids this matrix inversion. This allows us to handle the cases of
three and four impurities without much trouble. The resulting formulas for the dilatation operator are quite lengthy (see Appendix A), but 
their spectrum is surprisingly simple. 

{\sl
Our results suggest that for the class of operators considered, the Hamiltonian defined by the dilatation operator is integrable - it 
is just a set of oscillators. This is an example of an integrable dilatation operator, obtained by summing both planar and non-planar 
diagrams.
}

The operators we consider can be mapped to giant gravitons\cite{McGreevy:2000cw} 
in spacetime\cite{Balasubramanian:2001nh,Corley:2001zk,Berenstein:2004kk}. 
There is in fact already a known connection between the geometry of giant gravitons and harmonic 
oscillators\cite{Mikhailov:2000ya,Beasley:2002xv,Kinney:2005ej,Biswas:2006tj,Mandal:2006tk,jurgis}. 
Our work differs from these results in at least two important ways. Firstly,
we claim that the complete spectrum (not just the BPS spectrum!) has a connection to harmonic oscillators. Secondly, we have very 
good control over the set of operators we consider. Our operators are dual to a two giant system. Previous studies captured the full
set of BPS states and consequently were not able to distinguish (for example) giant graviton plus graviton from excited giant graviton.
Our study captures only states of the two giant system. Thus, it is rather natural to associate our oscillators with excitations
modes of a giant graviton. {\sl In the same way that there are oscillators in the worldsheet theory of a string describing the oscillation
modes of a string, we are describing the oscillators that would be present in a world volume description of giant gravitons describing
the oscillations modes of a giant graviton D3 brane.}

The problem of computing anomalous dimensions for operators with a large ${\cal R}$-charge has been considered before by
a number of authors. The restricted Schur polynomials we consider in this article are built by distributing ``impurities''
(given by $Y$s), in an operator built mainly from $Z$s. If we replace the $Y$s by words containing $O(\sqrt{N})$ letters
(which may be $Z$, $Y$ or other field or derivative of a field), these words are naturally identified with open 
strings\cite{Balasubramanian:2002sa,Sadri:2003mx,Berenstein:2003ah}.
In this case the dilatation operator reproduces the dynamics of open strings ending on a giant 
graviton\cite{Berenstein:2006qk,de Mello Koch:2007uu,de Mello Koch:2007uv,Bekker:2007ea}. 
The mixing of operators is highly constrained. Indeed, in \cite{de Mello Koch:2007uv,Bekker:2007ea} it 
was shown that operators which mix can differ
at most by moving one box around on the Young diagram labeling the operator. Another
interesting basis to consider is the Brauer basis\cite{Kimura:2007wy,Kimura:2009wy}. This basis is built using Brauer algebra projectors.
The structure constants of the Brauer algebra are $N$ dependent. There is an elegant construction of a class of BPS operators
\cite{yusuke} in which the natural $N$ dependence appearing in the definition of the operator\cite{heslop} is reproduced
by the Brauer algebra projectors\cite{yusuke}. Finally, another very natural approach to the problem, is to adopt a basis that
has sharp quantum numbers for the global symmetries of the theory\cite{Brown:2007xh,Brown:2008rr}. 
The action of the anomalous dimension operator in this sharp quantum number basis is very
similar to the action in the restricted Schur basis: again operators which mix can differ
at most by moving one box around on the Young diagram labeling the operator\cite{Brown:2008rs}.
For further related interesting work see \cite{tomyusuke,Huang:2010ne}.
Finally, for a rather general approach which correctly counts and constructs the weak coupling BPS operators see\cite{jurgis}.

We now conclude this introduction with a description of what is to follow.
In section 2 we will describe the action of the dilatation operator on restricted Schur polynomials. The main result of this section
is formula (\ref{final}) which gives a very explicit description of the action of the dilatation operator. In section 3 we will describe,
in broad terms,
the set of operators we consider and explain how the dynamics of
the large $N$ limit simplifies. In section 4 we explain how to construct the projectors
needed to evaluate the action of the dilatation operator. A very detailed description of the specific operators we consider is given in
section 5; we also describe a limit in which our system of two giants should be well described as a system of $D3$ brane giants plus open
strings. In this limit we see that the dilatation operator reduces to a lattice version of the second derivative, with the Young diagram
labels of the restricted Schur polynomials defining the lattice. In section 6 we present our numerical results and draw some general
conclusions from them in section 7. The explicit result for the dilatation operator is given in Appendix A; the intertwiners which enter
into the expression for the dilatation operator are described in Appendix B.

\section{Action of the Dilatation Operator}

{\sl In this section we will study the action of the one loop dilatation operator on restricted Schur polynomials
built using two complex adjoint scalars. The main result of this section is the surprisingly simple result (\ref{final})
for the action of the dilatation operator.}

{\vskip 0.2cm}

We will consider the action of the one loop dilatation operator in the $SU(2)$ sector\cite{Beisert:2003tq} of ${\cal N}=4$ super Yang
Mills theory
$$ 
D = - g_{\rm YM}^2 {\rm Tr}\,\big[ Y,Z\big]\big[ \partial_Y ,\partial_Z\big]
$$
on the restricted Schur polynomial 
$$
\chi_{R,(r,s)}(Z^{\otimes \, n},Y^{\otimes \, m})
=
{1\over n!m!}\sum_{\sigma\in S_{n+m}}{\rm Tr}_{(r,s)}(\Gamma_R(\sigma))
Z^{i_1}_{i_{\sigma(1)}}\cdots Z^{i_n}_{i_{\sigma(n)}}Y^{i_{n+1}}_{i_{\sigma(n+1)}}\cdots Y^{i_{n+m}}_{i_{\sigma(n+m)}}\, .
$$
The labels of our restricted Schur polynomial $\chi_{(R,(r,s))}$ are 
(i) $R$, which is a Young diagram with $n+m$ boxes or equivalently an irreducible representation of $S_{n+m}$,
(ii) $r$, which is a Young diagram with $n$ boxes or equivalently an irreducible representation of $S_n$ and
(iii) $s$ which is a Young diagram with $m$ boxes or equivalently an irreducible representation of $S_m$.
The notation ${\rm Tr}_{(r,s)}$ implies that one should only trace over the subspace carrying the irreducible 
representation\footnote{In general, because $(r,s)$ can be subduced more than once, we should include a multiplicity
index. We will not write this index explicitly in this article.}
 $(r,s)$ of $S_n\times S_m$ inside the carrier space for irreducible representation $R$ of $S_{n+m}$.
This trace is most concretely realized by including a projector $P_{R\to (r,s)}$ (from the carrier space of $R$ to 
the carrier space of $(r,s)$) and tracing over all of $R$.
A simple calculation yields\footnote{Our index conventions are $(YZ)^i_k = Y^i_j Z^j_k$.}  
{\small
\begin{eqnarray}
\label{basicreslt}
D \, \chi_{R,(r,s)}(Z^{\otimes \, n},Y^{\otimes \, m})=&&
{g_{\rm YM}^2\over (n-1)!(m-1)!}\sum_{\psi\in S_{n+m}}
\Tr_{(r,s)}\left(\Gamma_R ((n,n+1) \psi -\psi (n,n+1))\right)\times\cr
\times Z^{i_1}_{i_{\psi (1)}}&&
\cdots
Z^{i_{n-1}}_{i_{\psi (n-1)}}
(YZ-ZY)_{i_{\psi (n)}}^{i_n}\delta^{i_{n+1}}_{i_{\psi (n+1)}}
Y^{i_{n+2}}_{i_{\psi (n+2)}}\cdots Y^{i_{n+m}}_{i_{\psi (n+m)}}
\, .
\end{eqnarray}
}
The sum over $\psi$ runs only over permutations for which $\psi (n+1)=n+1$.
To perform this sum over $\psi$, write the sum over $S_{n+m}$ as a sum over cosets of the $S_{n+m-1}$ subgroup
obtained by keeping precisely those permutations that satisfy $\psi (n+1)=n+1$. The result follows immediately
from the reduction rule for Schur polynomials (see \cite{de Mello Koch:2004ws} and appendix C of \cite{de Mello Koch:2007uu})
{\small
\begin{eqnarray}
\nonumber
D \, \chi_{R,(r,s)}=&&
{g_{\rm YM}^2\over (n-1)!(m-1)!}\sum_{\psi\in S_{n+m-1}}\sum_{R'}c_{RR'}\,
\Tr_{(r,s)}\Big(\Gamma_R ((n,n+1))\Gamma_{R'}(\psi) \cr
&&-\Gamma_{R'}(\psi) \Gamma_R((n,n+1))\Big)
\, Z^{i_1}_{i_{\psi (1)}}
\cdots
Z^{i_{n-1}}_{i_{\psi (n-1)}}
(YZ-ZY)_{i_{\psi (n)}}^{i_n}
Y^{i_{n+2}}_{i_{\psi (n+2)}}\cdots Y^{i_{n+m}}_{i_{\psi (n+m)}}
\, .
\end{eqnarray}
}
The sum over $R'$ runs over all representations that can be subduced from $R$. Concretely, $R'$ runs over all Young
diagrams that can be obtained from $R$ by dropping a single box; $c_{RR'}$ is the 
weight\footnote{Recall that the weight of a box in row $i$ and column $j$ is $N - i + j$.} of the box that must be removed
from $R$ to obtain $R'$.
We will make use of the following notation for restricted characters
$$
\chi_{R,(r,s)}(\sigma )={\rm Tr}_{(r,s)}\Big(\Gamma_R(\sigma)\Big)={\rm Tr}\Big(P_{R\to (r,s)}\Gamma_R(\sigma)\Big)\, .
$$
Now, using the identity (bear in mind that $\psi (n+1)=n+1$)
$$
Z^{i_1}_{i_{\psi (1)}} \cdots Z^{i_{n-1}}_{i_{\psi (n-1)}} (YZ-ZY)_{i_{\psi (n)}}^{i_n}
Y^{i_{n+2}}_{i_{\psi (n+2)}} \cdots Y^{i_{n+m}}_{i_{\psi (n+m)}}
={\rm Tr}\left(\Big((n,n+1)\, \psi -\psi\,(n,n+1)\Big)Z^{\otimes n}Y^{\otimes m} \right)
$$
where
$$
{\rm Tr} (\sigma Z^{\otimes n}Y^{\otimes m})=
Z^{i_1}_{i_{\sigma(1)}}\cdots Z^{i_n}_{i_{\sigma(n)}}Y^{i_{n+1}}_{i_{\sigma(n+1)}}\cdots Y^{i_{n+m}}_{i_{\sigma(n+m)}}\, ,
$$
and (this identity is proved in \cite{Bhattacharyya:2008rc})
$$
{\rm Tr} (\sigma Z^{\otimes n}Y^{\otimes m})=\sum_{T,(t,u)}{d_T n! m!\over d_t d_u (n+m)!}\chi_{T,(t,u)}(\sigma^{-1})\chi_{T,(t,u)}(Z,Y)
$$
we obtain
$$
D\chi_{R,(r,s)}(Z,Y)=\sum_{T,(t,u)} M_{R,(r,s);T,(t,u)}\chi_{T,(t,u)}(Z,Y)\, ,
$$
{\small
$$
M_{R,(r,s);T,(t,u)}=g_{YM}^2\sum_{\psi\in S_{n+m-1}}\sum_{R'}
{c_{RR'} d_T n m\over d_t d_u (n+m)!}
\Tr_{(r,s)}\Big(\Gamma_R ((n,n+1))\Gamma_{R'}(\psi)-\Gamma_{R'}(\psi) \Gamma_R((n,n+1))\Big)\times
$$
$$
\times \chi_{T,(t,u)}(\psi^{-1}(n,n+1) - (n,n+1)\psi^{-1})\, .
$$
}
The sum over $\psi$ can be done by using the fundamental orthogonality relation
\begin{eqnarray}
\label{final}
M_{R,(r,s);T,(t,u)} &&= - g_{YM}^2\sum_{R'}{c_{RR'} d_T n m\over d_{R'} d_t d_u (n+m)}
\Tr\Big( \Big[ \Gamma_R((n,n+1)),P_{R\to (r,s)}\Big]I_{R'\, T'}\times \cr
&&\times \Big[\Gamma_T((n,n+1)),P_{T\to (t,u)}\Big] I_{T'\, R'}\Big)  \, .
\end{eqnarray}
The reader should consult Appendix B for a definition of the intertwiners $I_{R'\, T'}$.  
This expression for the one loop dilatation operator is exact in $N$. It is one of the key results of this article.

To obtain the spectrum of anomalous dimensions, we need to consider the action of the dilatation operator on normalized 
operators. The two point function for restricted Schur polynomials has been computed in \cite{Bhattacharyya:2008rb}
$$
\langle\chi_{R,(r,s)}(Z,Y)\chi_{T,(t,u)}(Z,Y)^\dagger\rangle =
\delta_{R,(r,s)\,T,(t,u)}f_R {{\rm hooks}_R\over {\rm hooks}_{r}\, {\rm hooks}_s}\, .
$$
In this expression $f_R$ is the product of the weights in Young diagram $R$ and ${\rm hooks}_R$ is the product of the hook 
lengths of Young diagram $R$. The normalized operators can be obtained from
$$
\chi_{R,(r,s)}(Z,Y)=\sqrt{f_R \, {\rm hooks}_R\over {\rm hooks}_r\, {\rm hooks}_s}O_{R,(r,s)}(Z,Y)\, .
$$
In terms of these normalized operators
$$
DO_{R,(r,s)}(Z,Y)=\sum_{T,(t,u)} N_{R,(r,s);T,(t,u)}O_{T,(t,u)}(Z,Y)
$$
{\small
$$
N_{R,(r,s);T,(t,u)}= - g_{YM}^2\sum_{R'}{c_{RR'} d_T n m\over d_{R'} d_t d_u (n+m)}
\sqrt{f_T \, {\rm hooks}_T\, {\rm hooks}_r \, {\rm hooks}_s \over f_R \, {\rm hooks}_R\, {\rm hooks}_t\, {\rm hooks}_u}\times
$$
$$
\times\Tr\Big(\Big[ \Gamma_R((n,n+1)),P_{R\to (r,s)}\Big]I_{R'\, T'}\Big[\Gamma_T((n,n+1)),P_{T\to (t,u)}\Big]I_{T'\, R'}\Big) \, .
$$
}
This last expression will be used later when we numerically study the spectrum of the dilatation operator.

\section{Excited Giant Graviton Bound States}

{\sl The goal of this section is to clearly define the class of operators being considered and to outline the approximations
that can be made in the large $N$ limit.}

{\vskip 0.2cm}

In this article we will study restricted Schur polynomials labeled by Young diagrams with at most two columns. The number of
$Z$s appearing is $\alpha N$ where $2-\alpha\equiv\zeta\ll 1$. The number of $Y$s appearing is fixed to be $O(1)$. These operators
are dual to giant gravitons that wrap an $S^3$ in the $S^5$ of the AdS$_5\times$S$^5$ background. Since the restricted Schur
polynomials furnish a suitable basis for the two giant system, we know that these operators capture all excitations (BPS and
non supersymmetric) of the two giant system. For a study of excitations of the single giant system using restricted Schur polynomials
see \cite{Koch:2010gp}. For a spacetime study of excitations of the single giant system using the Born-Infeld action see \cite{Das:2000st}.
The most important result from \cite{Das:2000st,Koch:2010gp} for us, is that all the deformations of the single threebrane giant graviton
that we consider, are supersymmetric.

The mixing of these operators with restricted Schur polynomials that have three columns (or more)
is suppressed by a factor of order ${1\over\sqrt{N}}$.
This factor arises from the normalization of the restricted Schur polynomials: the three column restricted Schur polynomials (with one 
short column) have a two point function which is smaller than the two column restricted Schur polynomials by a factor of order
${1\over N}$\cite{Koch:2010gp}. Thus, at large $N$ we can focus on the two column restricted Schur polynomials, which is a huge simplification.
The analog of the statement that for operators with a dimension of $O(1)$, different trace structures do not mix is: {\sl at
large $N$ restricted Schur polynomials $\chi_{R,(r,s)}$ with $R$ a Young diagram with $n$ columns, each of which has length of $O(N)$,
do not mix with operators $\chi_{R',(r',s')}$ that have $n'\ne n$ columns.}
The fact that the two column restricted Schur polynomials are a decoupled sector at large $N$ is to be expected. Indeed, at large $N$ these
operators correspond to a well defined stable semi-classical object in spacetime (the two giant system). We expect that $n$ column
restricted Schur polynomials are also a decoupled sector at large $N$ for the same reason.

\section{Simple Projectors}

{\sl When all Young diagrams labeling the restricted Schur polynomial have at most 2 columns, the projector $P_{R\to (r,s)}$
     simplifies dramatically. The goal of this section is to explain the simplification and exploit it to efficiently
     build the relevant projectors.}

{\vskip 0.2cm}

The projector $P_{R\to (r,s)}$ projects from representation $R$ of $S_{n+m}$ to representation $(r,s)$ of $S_n\times S_m$.
One issue, which complicates things considerably, is that representation $(r,s)$ can be subduced more than once when 
irreducible representation $R$ is decomposed into irreducible representations of the $S_n\times S_m$ subgroup. Giving a
general rule to specify precisely how this multiplicity is resolved is nontrivial. For the operators we consider in this
article, this problem does not arise. As soon as a system of three or more gaint gravitons are considered, it will be necessary
to deal with this issue. Fortunately a well defined approach to resolving these multiplicities has been outlined 
in \cite{Kimura:2008wy}. Basically, \cite{Kimura:2008wy} considers elements in the group algebra $CS_{n+m}$ which are 
invariant under conjugation by $CS_n\times CS_m$. The Cartan subalgebra of these elements are the natural generalization of the 
Jucys-Murphy elements which define a Cartan subalgebra for $S_n$\cite{Okunkov}. 
The multiplicities will be labelled by the eigenvalues of this Cartan subalgebra\cite{Kimura:2008wy}. 
It would be very interesting to work out the details of this proposal in the context of multigiant systems.

Given $R$ we imagine removing some boxes; after removing these boxes one is left with $r$. The removed boxes are assembled to
produce $s$. If we specify both $R$ and the boxes that are to be removed to obtain $r$ we obtain every representation exactly
once. For example, consider all possible operators that can be constructed from the following $R$ by removing the three boxes
shown
$$
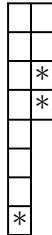
\young({\,}{\,},{\,}{\,},{\,}{*},{\,}{*},{\,},{\,},{\,},{*})
$$
For the boxes that are removed, we must respect edges that are joined, which means the two boxes removed from the short column must
remain stacked on top of each other. Thus, there are two possible irreducible representations $s$ that can be produced, implying
we can subduce two possible irreducible representations of $S_9 \times S_3$, by removing these three boxes
$$
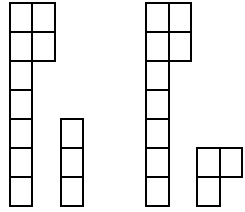
\young({\,}{\,},{\,}{\,},{\,},{\,},{\,},{\,},{\,})\, \young({\,},{\,},{\,})\qquad
\young({\,}{\,},{\,}{\,},{\,},{\,},{\,},{\,},{\,})\, \young({\,}{\,},{\,})
$$
Thus, in total we'd get the following $S_9 \times S_3$ irreducible representations subduced from the $R$ given above (sum over all
possible ways to remove boxes to get this result)
$$
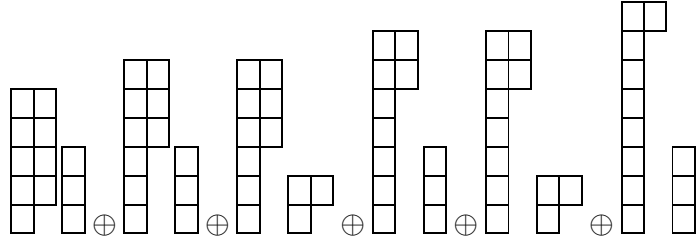
\young({\,}{\,},{\,}{\,},{\,}{\,},{\,}{\,},{\,})\,\young({\,},{\,},{\,})   \oplus
\young({\,}{\,},{\,}{\,},{\,}{\,},{\,},{\,},{\,})\,\young({\,},{\,},{\,}) \oplus
\young({\,}{\,},{\,}{\,},{\,}{\,},{\,},{\,},{\,})\,\young({\,}{\,},{\,})  \oplus
\young({\,}{\,},{\,}{\,},{\,},{\,},{\,},{\,},{\,})\,\young({\,},{\,},{\,}) \oplus
\young({\,}{\,},{\,}{\,},{\,},{\,},{\,},{\,},{\,})\,\young({\,}{\,},{\,})  \oplus
\young({\,}{\,},{\,},{\,},{\,},{\,},{\,},{\,},{\,})\,\young({\,},{\,},{\,})
$$
The dimensions of these representations are 42, 48, 96, 27, 54 and 8 respectively. This sums to give 275
which is indeed the dimension of $R$. 

To construct the actual projector, we need only build an operator which will assemble the removed boxes
in the correct way to produce $s$. We will give an example of how to construct this operator; the general
case should be clear. Lets start with the representation $R$ shown, removing the boxes indicated
$$
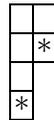
\young({\,}{\,},{\,}{*},{\,},{*})
$$
Our projector will act in the subspace spanned by the two sets of states
$$
|1\rangle =|\young({\,}{\,},{\,}{2},{\,},{1})\rangle\qquad
|2\rangle =|\young({\,}{\,},{\,}{1},{\,},{2})\rangle
$$
We are using a Young-Yamonouchi basis. Thus, each state given above could be any one of
$d_{\tiny\yng(2,1,1)}=3$ states, corresponding to the number of ways to complete the labels.
When acting in the subspace, the operator which organizes the boxes into representation $s$ is
$$ 
P_s={d_s\over m!}\sum_{\sigma\in S_m}\chi_s(\sigma)\Gamma_R(\sigma)\, .
$$
All that remains is to supply a formula for the action of $\Gamma_R(\sigma)$ (for $\sigma =1,(12)$) when
acting on the subspace spanned by $|1\rangle$ and $|2\rangle$.
The action of $\Gamma_R(\sigma)$ on any Young-Yamonouchi state is well known. For the states above
$$
\Gamma_R\left((12)\right)|1\rangle = -{1\over 3}|1\rangle +{\sqrt{8}\over 3}|2\rangle
$$
$$
\Gamma_R\left((12)\right)|2\rangle =  {1\over 3}|2\rangle +{\sqrt{8}\over 3}|1\rangle
$$
so that
$$
\Gamma_R\left( (12)\right)=-{1\over 3}|1\rangle\langle 1| +{\sqrt{8}\over 3}|2\rangle\langle 1|
+{1\over 3}|2\rangle\langle 2| +{\sqrt{8}\over 3}|1\rangle\langle 2|
$$
and
$$
\Gamma_R\left( 1\right)= |1\rangle\langle 1| + |2\rangle\langle 2| \, .
$$

\section{The Radial Direction}

{\sl In this section we describe a limit in which the dilatation operator simplifies significantly.
     There are two columns in the Young diagrams labeling the restricted Schur polynomials.
     When the first column contains $O(\sqrt{N})$ boxes more than the second, the dilatation operator simplifies 
     to a lattice realization of the second derivative. The Young diagram label itself defines the lattice.}

{\vskip 0.2cm}

\subsection{Three Impurities}

The three impurity operators are built using many $Z$s and three $Y$s. To specify these operators, we need to give the three
Young diagrams labeling the restricted Schur polynomial. The second Young diagram, $r$, (which specifies an irreducible representation
of $S_n$) is specified by stating the number of rows with two boxes ($=b_0$) and the number of rows with a single box ($=b_1$).
The third Young diagram label, $s$, (which specifies an irreducible representation of $S_3$) and the first Young diagram label, $R$, (which
specifies an irreducible representation of $S_{n+3}$) can now be built from $r$ by specifying which boxes in $R$ are to be removed to
obtain $r$ and how these boxes are to be organized into an $S_3$ irreducible representation. There are 6 possibilities
$$
\chi_A(b_0,b_1)=\chi_{\tiny \yng(2,2,2,2,2,1,1,1,1,1)\,\yng(2,2,2,2,2,1,1)\,\yng(1,1,1)}(Z,Y)
\qquad
\chi_B(b_0,b_1)=\chi_{\tiny \yng(2,2,2,2,2,1,1,1,1,1)\,\yng(2,2,2,2,1,1,1,1)\,\yng(1,1,1)}(Z,Y)
$$
$$
\chi_C(b_0,b_1)=\chi_{\tiny \yng(2,2,2,2,2,1,1,1,1,1)\,\yng(2,2,2,2,1,1,1,1)\,\yng(2,1)}(Z,Y)
\qquad
\chi_D(b_0,b_1)=\chi_{\tiny \yng(2,2,2,2,2,1,1,1,1,1)\,\yng(2,2,2,1,1,1,1,1,1)\,\yng(1,1,1)}(Z,Y)
$$
$$
\chi_E(b_0,b_1)=\chi_{\tiny \yng(2,2,2,2,2,1,1,1,1,1)\,\yng(2,2,2,1,1,1,1,1,1)\,\yng(2,1)}(Z,Y)
\qquad
\chi_F(b_0,b_1)=\chi_{\tiny \yng(2,2,2,2,2,1,1,1,1,1)\,\yng(2,2,1,1,1,1,1,1,1,1)\,\yng(1,1,1)}(Z,Y)
$$
The corresponding normalized operators are denoted using the capital letter $O$. In view of the discussion of section 3, we know that
$b_0$ is $O(N)$ and $b_1$ ranges from 0 or 1 to $O(N)$. The action of the dilatation operator is given in Appendix A.

The ${\cal R}$-charge of an operator in the field theory maps into the angular momentum of the dual string
theory state. Thanks to the Myers effect\cite{Myers:1999ps}
the angular momentum of the string theory state determines its size. Identifying the two columns of the Young
diagrams with the two threebranes, the number of boxes in each column determines the angular momentum and hence the size of each
threebrane. In the limit that $N-b_0=O(N)$, $b_0=O(N)$ and $b_1=O(\sqrt{N})$ we have non-maximal giants which are separated 
by a distance of $O(1)$ in string units. In this limit, we expect the dynamics to simplify. Indeed, the system should be described
by two $D3$ brane giant gravitons with open strings stretching between then. The action of the dilatation
operator becomes
$$
DO_A(b_0,b_1) =g_{YM}^2\left(N-b_0\right)\times O\left({1\over b_1}\right)
$$
\begin{eqnarray}
\nonumber
DO_B(b_0,b_1) =&&-{4\over 3}g_{YM}^2\left(N-b_0\right)\left[
O_B(b_0+1,b_1-2)-2\,O_B(b_0,b_1) + O_B(b_0-1,b_1+2) \right]\cr
&&+{2\sqrt {2}\over 3}g_{YM}^2\left(N-b_0\right)\left[O_C(b_0+1,b_1-2)
-2O_C(b_0+1,b_1)+O_C(b_0-1,b_1+2)\right]
\end{eqnarray}
\begin{eqnarray}
\nonumber
DO_C(b_0,b_1) =&&{2\sqrt {2}\over 3}g_{YM}^2(N-b_0)\left[
 O_B(b_0+1,b_1-2)-2O_B(b_0,b_1)+O_B(b_0-1,b_1+2)\right]\cr
&&-{2\over 3}g_{YM}^2(N-b_0)\left[ O_C(b_0+1,b_1-2)-2O_C(b_0,b_1)+O_C(b_0-1,b_1+2)\right]
\end{eqnarray}
\begin{eqnarray}
\nonumber
DO_D(b_0,b_1) =&&-{4\over 3}g_{YM}^2(N-b_0)\left[
O_D(b_0+1,b_1-2)-2O_D(b_0,b_1)+O_D(b_0-1,b_1+2)\right]\cr
&&+{2\sqrt {2}\over 3}g_{YM}^2(N-b_0)\left[
O_E(b_0+1,b_1-2)-2O_E(b_0,b_1)+O_E(b_0-1,b_1+2)\right]
\end{eqnarray}
\begin{eqnarray}
\nonumber
DO_E(b_0,b_1) = &&{2\sqrt {2}\over 3}g_{YM}^2(N-b_0)\left[
O_D(b_0+1,b_1-2)-2O_D(b_0,b_1)+O_D(b_0-1,b_1+2)\right]\cr
&&-{2\over 3}g_{YM}^2(N-b_0)\left[
O_E(b_0+1,b_1-2)-2O_E(b_0,b_1)+O_E(b_0-1,b_1+2)\right]
\end{eqnarray}
$$
DO_F(b_0,b_1) =g_{YM}^2\left(N-b_0\right)\times O\left({1\over b_1}\right)
$$
These results have a natural interpretation. Notice that there are four operators for which
the $S_m$ representation is the totally antisymmetric representation. We will see that there are also four operators for which 
the corresponding states remain supersymmetric; this agreement between the number of operators for which the $S_m$ representation
is the totally antisymmetric representation and the number of supersymmetric states, holds in general for the
two giant system. Looking at the labels, it is natural to interpret $O_A(b_0,b_1)$ as a state 
in which we deform only the larger threebrane. Recall from section 3, that deforming a single threebrane gives us a supersymmetric state 
so it seems natural for $O_A(b_0,b_1)$ to remain supersymmetric. Similarly, $O_F(b_0,b_1)$ can be interpreted as a state 
in which we deform only the smaller threebrane and a similar comment can be made. The fact that the combinations 
$O_B(b_0,b_1)+\sqrt{2}O_C(b_0,b_1)$ and $O_D(b_0,b_1)+\sqrt{2}O_E(b_0,b_1)$ are annihilated by $D$ implies that there are 
another two supersymmetric ways to deform the pair of threebranes. Finally, notice that if we set 
$O_B(b_0,b_1)-O_C(b_0,b_1)/\sqrt{2}\equiv O_{B-C}(b_0,b_1)$ and $O_D(b_0,b_1)-O_E(b_0,b_1)/\sqrt{2}\equiv O_{D-E}(b_0,b_1)$ 
we have
\begin{eqnarray}
\nonumber
DO_{B-C}(b_0,b_1) = &&-2g_{YM}^2(N-b_0)\left[
O_{B-C}(b_0+1,b_1-2)-2O_{B-C}(b_0,b_1)+O_{B-C}(b_0-1,b_1+2)\right]\cr
DO_{D-E}(b_0,b_1) = &&-2g_{YM}^2(N-b_0)\left[
O_{D-E}(b_0+1,b_1-2)-2O_{D-E}(b_0,b_1)+O_{D-E}(b_0-1,b_1+2)\right]
\end{eqnarray}
The right hand side again is a discretization of the second derivative. {\sl It is the Young diagram
itself that is defining the lattice.} After recalling that the number of boxes in each column sets the angular momentum 
and hence the radius\footnote{The giant graviton threebrane wraps an S$^3$ of a given radius. It is the radius of this
S$^3$ that we call the ``radius of the threebrane''.} of the corresponding threebrane, its clear that the radius of the giant graviton together with
local physics in this radial direction has emerged. 

\subsection{Four Impurities}

For the case of four impurities there are nine possible operators that we can define 
$$
\chi_A(b_0,b_1)=\chi_{\tiny \yng(2,2,2,2,2,2,1,1,1,1,1)\, \yng(2,2,2,2,2,2,1)\, \yng(1,1,1,1)}(Z,Y)
\qquad
\chi_B(b_0,b_1)=\chi_{\tiny \yng(2,2,2,2,2,2,1,1,1,1,1)\, \yng(2,2,2,2,2,1,1,1)\, \yng(1,1,1,1)}(Z,Y)
$$
$$
\chi_C(b_0,b_1)=\chi_{\tiny \yng(2,2,2,2,2,2,1,1,1,1,1)\,\yng(2,2,2,2,2,1,1,1)\, \yng(2,1,1)}(Z,Y)
\qquad
\chi_D(b_0,b_1)=\chi_{\tiny \yng(2,2,2,2,2,2,1,1,1,1,1)\, \yng(2,2,2,2,1,1,1,1,1)\, \yng(1,1,1,1)}(Z,Y)
$$
$$
\chi_E(b_0,b_1)=\chi_{\tiny \yng(2,2,2,2,2,2,1,1,1,1,1)\, \yng(2,2,2,2,1,1,1,1,1)\, \yng(2,1,1)}(Z,Y)
\qquad
\chi_F(b_0,b_1)=\chi_{\tiny \yng(2,2,2,2,2,2,1,1,1,1,1)\, \yng(2,2,2,2,1,1,1,1,1)\, \yng(2,2)}(Z,Y)
$$
$$
\chi_G(b_0,b_1)=\chi_{\tiny \yng(2,2,2,2,2,2,1,1,1,1,1)\, \yng(2,2,2,1,1,1,1,1,1,1)\, \yng(1,1,1,1)}(Z,Y)
\qquad
\chi_H(b_0,b_1)=\chi_{\tiny \yng(2,2,2,2,2,2,1,1,1,1,1)\, \yng(2,2,2,1,1,1,1,1,1,1)\, \yng(2,1,1)}(Z,Y)
$$
$$
\chi_I(b_0,b_1)=\chi_{\tiny \yng(2,2,2,2,2,2,1,1,1,1,1)\, \yng(2,2,1,1,1,1,1,1,1,1,1)\, \yng(1,1,1,1)}(Z,Y)
$$
Again, the corresponding normalized operators are denoted using the capital letter $O$. The action of the dilatation operator
is given in Appendix A.

In the limit that $N-b_0=O(N)$, $b_0=O(N)$ and $b_1=O(\sqrt{N})$ the dynamics again simplifies. The action of the dilatation
operator becomes
$$
DO_A(b_0,b_1)=(N-b_0)g_{YM}^2\times O\left({1\over b_1}\right)
$$
\begin{eqnarray}
\nonumber
DO_B(b_0,b_1) =&&-{3\over 2}g_{YM}^2\left(N-b_0\right)\left[
O_B(b_0+1,b_1-2)-2\,O_B(b_0,b_1) + O_B(b_0-1,b_1+2) \right]\cr
&&+{\sqrt {3}\over 2}g_{YM}^2\left(N-b_0\right)\left[O_C(b_0+1,b_1-2)
-2O_C(b_0+1,b_1)+O_C(b_0-1,b_1+2)\right]
\end{eqnarray}
\begin{eqnarray}
\nonumber
DO_C(b_0,b_1) =&&{\sqrt {3}\over 2}g_{YM}^2(N-b_0)\left[
 O_B(b_0+1,b_1-2)-2O_B(b_0,b_1)+O_B(b_0-1,b_1+2)\right]\cr
&&-{1\over 2}g_{YM}^2(N-b_0)\left[ O_C(b_0+1,b_1-2)-2O_C(b_0,b_1)+O_C(b_0-1,b_1+2)\right]
\end{eqnarray}
\begin{eqnarray}
\nonumber
DO_D(b_0,b_1) =&&-2g_{YM}^2\left(N-b_0\right)\left[
O_D(b_0+1,b_1-2)-2\,O_D(b_0,b_1) + O_D(b_0-1,b_1+2) \right]\cr
&&+{2\over \sqrt{3}}g_{YM}^2\left(N-b_0\right)\left[O_E(b_0+1,b_1-2)
-2O_E(b_0+1,b_1)+O_E(b_0-1,b_1+2)\right]
\end{eqnarray}
\begin{eqnarray}
\nonumber
DO_E(b_0,b_1) =&&-2g_{YM}^2(N-b_0)\left[
 O_E(b_0+1,b_1-2)-2O_E(b_0,b_1)+O_E(b_0-1,b_1+2)\right]\cr
&&+{2\over \sqrt{3}}g_{YM}^2(N-b_0)\left[ O_D(b_0+1,b_1-2)-2O_D(b_0,b_1)+O_D(b_0-1,b_1+2)\right]\cr
&&+{2\sqrt{6}\over 3}g_{YM}^2(N-b_0)\left[ O_F(b_0+1,b_1-2)-2O_F(b_0,b_1)+O_F(b_0-1,b_1+2)\right]
\end{eqnarray}
\begin{eqnarray}
\nonumber
DO_F(b_0,b_1) =&&-2g_{YM}^2\left(N-b_0\right)\left[
O_F(b_0+1,b_1-2)-2\,O_F(b_0,b_1) + O_F(b_0-1,b_1+2) \right]\cr
&&+{2\sqrt{6}\over 3}g_{YM}^2\left(N-b_0\right)\left[O_E(b_0+1,b_1-2)
-2O_E(b_0+1,b_1)+O_E(b_0-1,b_1+2)\right]
\end{eqnarray}
\begin{eqnarray}
\nonumber
DO_G(b_0,b_1) =&&-{3\over 2}g_{YM}^2\left(N-b_0\right)\left[
O_G(b_0+1,b_1-2)-2\,O_G(b_0,b_1) + O_G(b_0-1,b_1+2) \right]\cr
&&+{\sqrt{3}\over 2}g_{YM}^2\left(N-b_0\right)\left[O_H(b_0+1,b_1-2)
-2O_H(b_0+1,b_1)+O_H(b_0-1,b_1+2)\right]
\end{eqnarray}
\begin{eqnarray}
\nonumber
DO_H(b_0,b_1) =&&-{1\over 2}g_{YM}^2\left(N-b_0\right)\left[
O_H(b_0+1,b_1-2)-2\,O_H(b_0,b_1) + O_H(b_0-1,b_1+2) \right]\cr
&&+{\sqrt{3}\over 2}g_{YM}^2\left(N-b_0\right)\left[O_G(b_0+1,b_1-2)
-2O_G(b_0+1,b_1)+O_G(b_0-1,b_1+2)\right]
\end{eqnarray}
$$
DO_I(b_0,b_1)=(N-b_0)g_{YM}^2\times O\left({1\over b_1}\right)
$$
We can again identify combinations of operators that are annihilated by $D$, that is, that are BPS. Apart from
$O_A(b_0,b_1)$ and $O_I(b_0,b_1)$ we have $O_B(b_0,b_1)+\sqrt{3}O_C(b_0,b_1)$,
$O_D(b_0,b_1)+\sqrt{3}O_E(b_0,b_1)+\sqrt{2}O_F(b_0,b_1)$ and $O_G(b_0,b_1)+\sqrt{3}O_H(b_0,b_1)$. Notice that all of
the BPS operators from this section and the last can be written as
$$
O_{\rm BPS}(R,r)=\sum_s \sqrt{d_s} O_{R,(r,s)}(b_0,b_1)
$$
where $d_s$ is the dimension of the irreducible representation $s$ of the symmetric group. Finally, notice that if we set 
$\sqrt{3}O_B(b_0,b_1)-O_C(b_0,b_1)\equiv O_{B-C}(b_0,b_1)$,  
$\sqrt{2}O_D(b_0,b_1)-O_F(b_0,b_1)\equiv O_{D-F}(b_0,b_1)$,
$O_D(b_0,b_1)-\sqrt{3}O_E(b_0,b_1)+\sqrt{2}O_F(b_0,b_1)\equiv O_{DF-E}(b_0,b_1)$ and
$\sqrt{3}O_G(b_0,b_1)-O_H(b_0,b_1)\equiv O_{G-H}(b_0,b_1)$, we have
{\small
\begin{eqnarray}
\nonumber
DO_{B-C}(b_0,b_1) = &&-2g_{YM}^2(N-b_0)\left[
O_{B-C}(b_0+1,b_1-2)-2O_{B-C}(b_0,b_1)+O_{B-C}(b_0-1,b_1+2)\right]\cr
DO_{D-F}(b_0,b_1) = &&-2g_{YM}^2(N-b_0)\left[
O_{D-F}(b_0+1,b_1-2)-2O_{D-F}(b_0,b_1)+O_{D-F}(b_0-1,b_1+2)\right]\cr
DO_{DF-E}(b_0,b_1) = &&-4g_{YM}^2(N-b_0)\left[
O_{DF-E}(b_0+1,b_1-2)-2O_{DF-E}(b_0,b_1)+O_{DF-E}(b_0-1,b_1+2)\right]\cr
DO_{G-H}(b_0,b_1) = &&-2g_{YM}^2(N-b_0)\left[
O_{G-H}(b_0+1,b_1-2)-2O_{G-H}(b_0,b_1)+O_{G-H}(b_0-1,b_1+2)\right]
\end{eqnarray}
}
The right hand side is again a discretization of the second derivative.

\section{Numerical Results}

{\sl In this section we describe the result of numerically diagonalizing the dilatation operator.}

{\vskip 0.2cm}

When setting up a numerical computation of the spectrum of the anomalous dimension operator, we need to specify the
maximum value for the difference between the number of boxes in the long column and the number of boxes in the short
column. Given this value, denoted $a_{\rm max}$, we are able to determine how many operators participate in our problem
and we are able to describe the resulting spectrum rather explicitly. We will focus on the case of even $a_{\rm max}$.
In this case the difference between the number of boxes in the long column and the number of boxes in the short
column is always an even number.

\subsection{Two Impurities}

For a given value of $a_{\rm max}$ there are $2+2a_{\rm max}$ states in total. There are ${3\over 2}a_{\rm max}+1$ zero
eigenvalues (corresponding to supersymmetric states). The remaining eigenvalues are
$$
\lambda_i =8g_{YM}^2 i\qquad i=1,2,\cdots,{a_{\rm max}\over 2}+1\, .
$$

\subsection{Three Impurities}

For a given value of $a_{\rm max}$ there are $1+3a_{\rm max}$ states in total. There are $2 a_{\rm max}$ zero
eigenvalues (corresponding to supersymmetric states). The remaining eigenvalues are
$$
\lambda_i =8g_{YM}^2 i\qquad i=1,2,\cdots,{a_{\rm max}\over 2}\, ,
$$
each with a degeneracy of two and a single maximum eigenvalue $\lambda=4a_{\rm max}g_{YM}^2+8g_{YM}^2$. 
This degeneracy almost certainly indicates a symmetry enhancement in the large $N$ limit.

\subsection{Four Impurities}

For a given value of $a_{\rm max}$ there are $1 +{9\over 2}a_{\rm max}$ states in total.
There are ${5\over 2}a_{\rm max}-1$ zero eigenvalues. The eigenvalues are again evenly spaced with a 
level spacing of $8g_{YM}^2$ and they are again degenerate. The low lying eigenvalues
$$
\lambda_i = 8 g_{YM}^2 i\qquad i=1,2,\cdots,{a_{\rm max}\over 2}\, ,
$$
have a degeneracy which alternates between 3 and 4. Thus, we find three eigenvalues $\lambda =8g_{YM}^2$, followed
by four eigenvalues $\lambda=16g_{YM}^2$, followed by three eigenvalues $\lambda=24g_{YM}^2$,
followed by four eigenvalues $\lambda = 32g_{YM}^2$ and so on. 

If $a_{\rm max}$ is a multiple of 4, the larger eigenvalues are given by $\lambda=4a_{\rm max}g_{YM}^2 +8g_{YM}^2$,
$\lambda=4a_{\rm max}g_{YM}^2 + 16g_{YM}^2$ and 
$$
\lambda_i=4a_{\rm max}g_{YM}^2 + 16g_{YM}^2 +16ig_{YM}^2\qquad i=1,2,\cdots,{a_{\rm max}\over 4}\, ;
$$
all of these eigenvalues on the last line above are non-degenerate.

If $a_{\rm max}$ (which by assumption is even) is not a multiple of 4, the larger eigenvalues are given by 
$\lambda=4a_{\rm max}g_{YM}^2 +8g_{YM}^2$ with a degeneracy of 2 and 
$$
\lambda_i=4a_{\rm max}g_{YM}^2 + 16g_{YM}^2 +16ig_{YM}^2\qquad i=1,2,\cdots,{a_{\rm max}+2\over 4}\, ;
$$
all of these eigenvalues on the last line above are non-degenerate.

Once again, the degeneracies observed almost certainly indicate a symmetry enhancement in the large $N$ limit.

\section{Discussion}

In this article we have computed the one loop anomalous dimension of an operator built from $O(N)$ $Z$s and 3 or 4
$Y$ ``impurities''. What lessons can be learnt from these results, together with the results of \cite{Koch:2010gp}, 
which dealt with the case of 2 impurities? Before we start the discussion, it is useful to recall the structure
of the operators which participate in the case of two impurities
$$
\chi_A(b_0,b_1)=\chi_{\tiny \yng(2,2,2,2,2,1,1,1,1)\,\yng(2,2,2,2,2,1,1)\,\yng(1,1)}(Z,Y)
\qquad
\chi_B(b_0,b_1)=\chi_{\tiny \yng(2,2,2,2,2,1,1,1,1)\,\yng(2,2,2,1,1,1,1,1,1)\,\yng(1,1)}(Z,Y)
$$
$$
\chi_D(b_0,b_1)=\chi_{\tiny \yng(2,2,2,2,2,1,1,1,1)\,\yng(2,2,2,2,1,1,1,1)\,\yng(2)}(Z,Y)
\qquad
\chi_E(b_0,b_1)=\chi_{\tiny \yng(2,2,2,2,2,1,1,1,1)\,\yng(2,2,2,2,1,1,1,1)\,\yng(1,1)}(Z,Y)
$$

Firstly, the actual result for the one loop dilation operator (see Appendix
A) is rather complicated. This is to be expected - it was obtained by summing 
a huge class of Feynman diagrams - much more than just the planar diagrams.
On the other hand, the spectra of anomalous dimensions obtained are rather simple.
To obtain a numerically tractable problem, we have been forced to keep the value
of $a_{\rm max}$ finite (recall that this parameter measures the maximum value 
for the difference between the number of boxes in the long column and the number 
of boxes in the short column). In the large $N$ limit\footnote{$\zeta$ was defined in section 3.} 
$a_{\rm max}=\zeta N$ goes to
infinity. Our discussion assumes we are working in this
$a_{\rm max}\to\infty$ limit.

We have restricted ourselves to a study of less than five impurities. This restriction is
not imposed because our methods break down for five or more impurities, but only because the
details of writing down the projection operators and evaluating the dilatation operator
becomes increasingly complicated as the number of impurities is increased. In particular, there
is no problem in principle with taking $O(N)$ impurities. The case of no impurities or one
impurity are simple to handle analytically - all of these operators are annihilated by the one loop 
dilatation operator\cite{Koch:2010gp}. 

For the case of two impurities there are three times as many zero eigenvalue states
as there positive eigenvalue states. There are ${a_{\rm max}\over 2}$ positive eigenvalue
states with a constant energy level spacing of $8g_{YM}^2$. 
Thus, it is natural to associate an oscillator with a set of $\sim{a_{\rm max}\over 2}$ states.
With this assumption, the dilatation operator acting on the two impurity
operators gives the spectrum of three harmonic oscillators with
a level spacing of zero and a single harmonic oscillator with a level spacing of $8g_{YM}^2$.
Looking at the two impurity operators given above, we see that there are three operators
with impurities in the antisymmetric representation (${\tiny \yng(1,1)}$) and one operator with 
the impurities in the symmetric representation (${\tiny \yng(2)}$). 

For the case of three impurities, the dilatation operator gives the spectrum of four harmonic 
oscillators with a level spacing of zero and two harmonic oscillators with a level spacing of 
$8g_{YM}^2$; each oscillator again has $\sim{a_{\rm max}\over 2}$ states. Looking at the three 
impurity operators given in section 5.1, we see that there are four operators (A,B,D and F)
with impurities in the antisymmetric representation (${\tiny \yng(1,1,1)}$) and two operators (C and E)
with the impurities in the ${\tiny \yng(2,1)}$ representation. 

For the case of four impurities, recall that we had an interesting degeneracy structure - the degeneracy
alternates between three degenerate states and four degenerate states. This is naturally explained as
three oscillators with a level spacing of $8g_{YM}^2$ and a fourth with a level spacing of $16g_{YM}^2$;
with this interpretation each oscillator again has $\sim{a_{\rm max}\over 2}$ states. Thus, for the case 
of four impurities, the dilatation operator gives the spectrum of five harmonic oscillators with a level 
spacing of zero, three harmonic oscillator with a level spacing of $8g_{YM}^2$ and one with a spacing of
$16g_{YM}^2$. Looking at the four impurity operators given in section 5.2, we see that there are five operators 
(A,B,D,G and I) with impurities in the antisymmetric representation (${\tiny \yng(1,1,1,1)}$), three operators 
(C,E and H) with the impurities in the ${\tiny \yng(2,1,1)}$ representation and one (F) with impurities in the
${\tiny \yng(2,2)}$ representation\footnote{Note that by looking at the representation that organizes the impurities we
have been able to read off the frequencies of the harmonic oscillators appearing. This is where it ends; in particular
we are {\it not} claiming that $O_F$ are the operators corresponding to the frequency $16g_{YM}^2$ operators! Operators
with a good scaling dimension are a complicated linear combination of the various possible $O$s.}. 

It is rather easy to guess the result for a general number of impurities. If the number of impurities is even
$=2n$ we expect to obtain a set of oscillators with frequency $\omega_i$ and degeneracy $d_i$ given by
$$\omega_i =8ig_{YM}^2,\qquad d_i=2(n-i)+1,\qquad i=0,1,...,n\, .$$
If the number of impurities is odd $=2n+1$
we expect to obtain a set of oscillators with frequency $\omega_i$ and degeneracy $d_i$ given by
$$\omega_i =8ig_{YM}^2,\qquad d_i=2(n-i+1),\qquad i=0,1,...,n\, .$$
This conjecture passes a simple counting test: $\sum_i d_i$ is equal to the number of restricted Schur
polynomials that can be defined. Further, the degeneracies $d_i$ match the number of each type of
oscillator that can be defined: $d_i$ is equal to the number of operators which have the impurities
organized into a Young diagram with $i$ boxes in the short column.

A beautiful simple picture is emerging from the rather complicated formulas obtained for the dilatation operator: the
dilatation operator is equivalent to a set of harmonic oscillators. For each type of operator there is a single oscillator
and the frequency of the oscillator is determined by the representation which organizes the impurities. Since a set
of harmonic oscillators is an integrable system, this system we have studied here is an example of an integrable
dilatation operator, obtained by summing planar and non-planar diagrams.

${\cal N}=4$ super Yang-Mills theory has an $SU(4)$ ${\cal R}$-symmetry. A $U(2)$ subgroup of the ${\cal R}$-symmetry
can be used to rotate the $Y$s and $Z$s into each other. Thus, although the operators we have studied do not treat the
$Y$s and $Z$s symmetrically, a more complete study working with the complete set of two column restricted Schur polynomials
would yield a description in which $Z$ and $Y$ appear on an equal footing. If we are to recover this symmetric description
its clear that the oscillators which emerge for each particular impurity configuration must themselves have frequencies
which are multiples of $8g_{YM}^2$ - exactly as we have found. Exploring this further we should be able to show that the
spectrum is filling out $U(2)$ multiplets. By including more species of impurities we should also be able to see more
of the expected global symmetry group.

What is the AdS/CFT dual interpretation of our results? The operators we have considered are dual to giant gravitons.
A connection between the geometry of giant gravitons and harmonic oscillators was already uncovered in \cite{Biswas:2006tj,Mandal:2006tk,jurgis}. 
This work quantizes the moduli space of Mikhailov's giant gravitons. Consequently one is capturing a huge space of
states. It is this huge space of states that connects to harmonic oscillators. Our study has focused on a two
giant system. Consequently, we know that the oscillators that we have captured are associated to this two giant
system and excitations of it. We have thus arrived at a slightly more refined statement of how the harmonic oscillator
enters. The critical reader might question whether our set of operators includes excitations 
corresponding (for example) to the two giant system plus a graviton. This would seem to be a small perturbation
of the two giant system that is not an excitation of it - the graviton is an excitation of spacetime. We do
not have such excitations among our states: these states correspond to operators with a 
small third column, which have decoupled at large $N$. 
In much the same way that by quantizing the possible excitation modes of a string one obtains a set of oscillators, its
natural to think that our oscillators arise from the quantization of the possible excitation modes of a giant graviton.

We have described a limit (the first column of the Young diagram contains $O(\sqrt{N})$ boxes 
more than the second) in which the dilatation operator simplifies to a lattice realization of the second derivative.
{\sl It is the Young diagram itself that is defining the lattice.} 
After recalling that the number of boxes in each column sets the angular momentum 
and hence the radius of the corresponding threebrane, its clear that the radius of the giant graviton together with
local physics in this radial direction has emerged. Notice that for BMN loops the number of lattice sites is $O(\sqrt{N})$;
for the operators we have studied here the number of lattice sites is $O(N)$.

One can contemplate a number of tests for our proposal. With a thorough understanding of the vibrational modes of the giant
graviton system, one could imagine realizing definite classical membrane geometries by considering coherent states of the oscillators.
One might then compare the energy of these states with the volume of the membrane times the membrane tension. Although naively
the field theory and gravity computations are never simultaneously valid, one might hope that for operators which are close to BPS,
the perturbative result might agree with the strong coupling answer (see \cite{Berenstein:2003ah} for a relevant discussion).

In addition to questions we pointed out above, there are a number of clear directions for further study. 
Given the simplicity of our results, it should be possible to construct an analytic solution. 
This is under investigation\cite{bhw}.
It would also be interesting to understand how our results are modified at higher loops.
One could also consider the case of $n>2$ column restricted Schur polynomials and more species of impurities. 
A much more general question would be to ask when (and how) simple systems are expected to emerge from multimatrix models.
For a single matrix model, it is well known that the planar limit is nicely captured by the dynamics of $N$ non-interacting 
non-relativistic fermions in an external potential.
In this paper we have argued that the large $N$ limit of a class of operators dual to giant gravitons is captured by a collection
of harmonic oscillators.
Presumably every semiclassical object in spacetime (string, giant graviton, black hole, etc) is associated with the emergence of
a simple system in the large $N$ limit of the corresponding class of operators in the field theory. 
Can we make this connection sharper and more useful?

{\vskip 1.0cm}

\noindent
{\it Acknowledgements:}
We would like to thank Tom Brown, Warren Carlson, Norman Ives, Yusuke Kimura, Hai Lin, Dimitri Polyakov, 
Sanjaye Ramgoolam, Peter Roenne, Stephanie Smith and Michael Stephanou for pleasant discussions and/or 
helpful correspondence. This work is based upon research supported by the South African Research Chairs
Initiative of the Department of Science and Technology and National Research Foundation.
Any opinion, findings and conclusions or recommendations expressed in this material
are those of the authors and therefore the NRF and DST do not accept any liability
with regard thereto.

\appendix

\section{Dilatation Operator for Three or Four impurities}

The dilatation operator for the case of two impurities has been given in \cite{Koch:2010gp}. In what follows
$$
DO=g_{YM}^2\hat{D}O\, .
$$

\subsection{Three Impurities}

{\small

$$ \hat{D}O_A(b_0,b_1) =\sqrt{\left(N-b_0-b_1-2\right)\left(N-b_0+1\right)}\left[
4\,b_1\,\sqrt{{\frac{b_1+4}{b_1+2}}}{1\over \left(b_1+2\right) \left( b_1+3 \right)}
O_B(b_0,b_1)\right.
$$
$$
-2\,\sqrt {{\frac {b_1+4}{b_1+2}}}\sqrt {2}{1\over \left( b_1+2 \right)}
O_C(b_0,b_1)
+8\,\sqrt{\left(b_1+4\right)\left(b_1+1\right)\over\left(b_1+2\right)\left(b_1+3\right)}
{\frac{1}{\left(b_1+3 \right)\left( b_1+2 \right)}} O_D(b_0-1,b_1+2)
$$
$$
\left. +2\,\sqrt{\left(b_1+4\right)\left(b_1+1\right)\over \left(b_1+3\right)\left(b_1+2\right)}{\frac {\sqrt {2}}
{ \left( b_1+3 \right)\left( b_1+2 \right)}}O_E(b_0-1,b_1+2)\right]
+(N-b_0-b_1-2)\left[{\frac {12}{ \left( b_1+2 \right)  \left( b_1+3 \right) }}O_A(b_0,b_1)\right.
$$
$$
\left.
-4\,{\frac {\sqrt { \left( b_1+1 \right)  \left( b_1+3 \right) } \left( b_1+5 \right) }{ \left( b_1+3 \right) ^{2} \left( b_1+2 \right) }}O_B(b_0-1,b_1+2)
+2\,{\frac {\sqrt { \left( b_1+1 \right)  \left( b_1+3 \right) }\sqrt {2}}{ \left( b_1+3 \right) ^{2}}}O_C(b_0-1,b_1+2)\right]
$$
{\vskip 1cm}
$$ \hat{D}O_B(b_0,b_1) =\sqrt{\left(N-b_0-b_1-1\right)\left(N-b_0\right)}\left[
-{4\over 3}\, \sqrt {{\frac { \left( b_1+2 \right)  \left( b_1-1 \right) }{b_1\, \left( b_1+1 \right) }}}
{\left( b_1-2 \right)  \left( b_1+3 \right) \over b_1\left( b_1+1 \right)}O_B(b_0+1,b_1-2)\right.
$$
$$
+{2\over 3}\,{b_1+3\over b1}\sqrt{{\frac{\left(b_1+2\right)\left(b_1-1\right)}{\left(b_1+1\right)b_1}}}\sqrt {2}O_C(b_0+1,b_1-2)
-{32\over 3}\,{{b_1}^{2}+2\,b_1-3 \over b_1(b_1+1)(b_1+2)^2} \sqrt {{\frac {b_1+2}{b_1}}}O_D(b_0,b_1)
$$
$$
\left.
-{2\sqrt{2}\over 3}\sqrt {b_1+2\over b_1}{\frac {(b_1+3) (3b_1-2)}{b_1(b_1+2)(b_1+1)}}O_E(b_0,b_1)
+8\,\sqrt {\left( b_1+3 \right)b_1\over \left( b_1+2 \right)\left( b_1+1 \right)}{\frac {1}{\left(b_1+1\right)\left(b_1+2\right)}}O_F(b_0-1,b_1+2)
\right]
$$
$$
+\sqrt{\left(N-b_0-b_1-2\right)\left(N-b_0+1\right)}\left[
{2\over 3}\sqrt{\left(b_1+4\right)\left(b_1+1\right)\over \left(b_1+2\right)\left(b_1+3\right)}
{\frac {\sqrt{2}b_1}{\left(b_1+3\right)}}O_C(b_0-1,b_1+2)\right.
$$
$$
-{4\over 3}\,\sqrt { \left( b_1+4 \right)\left( b_1+1 \right)\over\left( b_1+3 \right)\left( b_1+2 \right)}
{\frac { \left( b_1+5 \right) b_1}{ \left( b_1+3 \right)\left( b_1+2 \right)}}O_B(b_0-1,b_1+2)
$$
$$
\left.
+4\,\sqrt {b_1+4\over b_1+2}{\frac {b_1}{\left( b_1+3 \right)\left( b_1+2 \right)}}O_A(b_0,b_1)\right]
+\left( N-b_0-b_1-1 \right)\left[
-4\sqrt {{\frac {b_1-1}{b_1+1}}}{\left( b_1+3 \right)\over\left(b_1+1\right) b_1}O_A(b_0+1,b_1-2)\right.
$$
$$
+{4\over 3}\,{\frac{(b_1+3)(b_1^3+5b_1^2+8b_1-12)}{(b_1+1)b_1(b_1+2)^2}}O_B(b_0,b_1)
-{2\sqrt {2}\over 3}\,{\frac {(b_1^2+2b_1-4)(b_1+3)}{(b_1+1)(b_1+2)^2}}O_C(b_0,b_1)
$$
$$
\left.
-{8\over 3}\, \sqrt {{\frac {b_1+3}{b_1+1}}} 
{\left( b_1+4 \right) b_1 \over \left( b_1+2 \right)^2 \left( b_1+1 \right)}O_D(b_0-1,b_1+2)
+{4\over 3}\, \sqrt {2}\sqrt {{\frac {b_1+3}{b_1+1}}}{b_1\over \left( b_1+2 \right)^2}
O_E(b_0-1,b_1+2)\right]
$$
$$
+\left( N-b_0+1 \right) \left[
{4\over 3}\,{\frac { \left( b_1+4 \right) {b_1}^{2} }{ \left( b_1+3 \right)  \left( b_1+2 \right) ^{2}}}O_B(b_0,b_1)
+{8\over 3}\,{\frac{\sqrt { \left( b_1+1 \right)  \left( b_1+3 \right) }b_1\, \left( b_1+4 \right) }
{ \left( b_1+3 \right) ^{2} \left( b_1+2 \right) ^{2}}}O_D(b_0-1,b_1+2)\right.
$$
$$
\left.
-{2\over 3}\,{\frac{\sqrt {2} \left( b_1+4 \right) b_1}{ \left( b_1+2 \right) ^{2}}}
O_C(b_0,b_1)
+{2\over 3}\,{\frac{\sqrt {2}\sqrt { \left( b_1+1 \right)  \left( b_1+3 \right) }b_1\, \left( b_1+4 \right) }
{ \left( b_1+3 \right) ^{2} \left( b_1+2 \right) ^{2}}}O_E(b_0-1,b_1+2)\right]
$$
{\vskip 1cm}
$$ \hat{D}O_C(b_0,b_1) =\sqrt {\left(N-b_0-b_1-1\right)\left(N-b_0\right)}\left[
{2\sqrt {2}\over 3}\,  \sqrt {{\frac { \left( b_1+2 \right)  \left( b_1-1 \right) }{ \left( b_1+1 \right) b_1}}}
{\left( b_1-2 \right)\over b_1+1 }O_B(b_0+1,b_1-2)\right.
$$
$$
-{2\over 3}\,\sqrt {{\frac { \left( b_1+2 \right)  \left( b_1-1 \right) }{ \left( b_1+1 \right) b_1}}}O_C(b_0+1,b_1-2)
+{2\sqrt{2}\over 3}\sqrt{{\frac {b_1+2}{b_1}}}{(b_1-1)(3b_1+8) \over (b_1+1)(b_1+2)^2}O_D(b_0,b_1)
$$
$$
\left.
-{4\over 3}\sqrt{b_1+2\over b_1}{1\over (b_1+1)(b_1+2)}O_E(b_0,b_1)
+2\,{\frac{\sqrt{\left(b_1+2\right)\left(b_1+3\right)\left(b_1+1\right)b_1}\sqrt{2}}{\left( b_1+1 \right)^2\left(b_1+2\right)^2}}O_F(b_0-1,b_1+2)
\right]
$$
$$
+\sqrt {\left(N-b_0-b_1-2\right)\left(N-b_0+1\right)}\left[
-2\,{\frac {\sqrt { \left( b_1+4 \right)  \left( b_1+2 \right) }\sqrt {2}}{ \left( b_1+2 \right) ^{2}}}O_A(b_0,b_1)\right.
$$
$$
\left.
+{2\sqrt{2}\over 3}\,\sqrt{\left(b_1+4\right)\left(b_1+1\right)\over \left(b_1+3\right)\left(b_1+2\right)}
{\frac{\left(b_1+5\right)}{\left(b_1+2\right)}}
O_B(b_0-1,b_1+2)
-{2\over 3}\,\sqrt{\left(b_1+4\right)\left(b_1+1\right)\over \left(b_1+2\right)\left(b_1+3\right)}
O_C(b_0-1,b_1+2)\right]
$$
$$
+\left( N-b_0-b_1-1 \right)
\left[2\,\sqrt {{\frac {b_1-1}{b_1+1}}}\sqrt {2}   {1\over b_1+1}O_A(b_0+1,b_1-2)\right.
$$
$$
-{2\sqrt{2}\over 3}\,{\frac {(b_1^2+2b_1-4)(b_1+3)}{(b_1+1)(b_1+2)^2}}O_B(b_0,b_1)
+{2\over 3}{\frac{b_1(b_1^2+2b_1-1)}{(b_1+1)(b_1+2)^2}}O_C(b_0,b_1)
$$
$$
\left.
-{2\over 3}\, \sqrt {2}\sqrt {{\frac {b_1+3}{b_1+1}}}
{ \left( b_1+4 \right) b_1 \over \left( b_1+2 \right)^2 \left( b_1+1 \right)}O_D(b_0-1,b_1+2)
+{2\over 3}\, \sqrt {{\frac {b_1+3}{b_1+1}}}{b_1\over \left( b_1+2 \right)^2}
O_E(b_0-1,b_1+2)\right]
$$
$$
+\left( N-b_0+1 \right)\left[
-{2\over 3}\,{\frac {  \sqrt {2} \left( b_1+4 \right) b_1}{ \left( b_1+2 \right) ^{2}}}O_B(b_0,b_1)
+{2\over 3}\,{\frac {  \left( b_1+4 \right)  \left( b_1+3 \right) }{ \left( b_1+2 \right) ^{2}}}O_C(b_0,b_1)\right.
$$
$$
\left.
-{4\over 3}\,{\frac {\sqrt {2}\sqrt { \left( b_1+1 \right)  \left( b_1+3 \right) } \left( b_1+4 \right) }{ \left( b_1+3 \right)  \left( b_1+2 \right) ^{2}}}O_D(b_0-1,b_1+2)
-{2\over 3}\,{\frac {\sqrt { \left( b_1+1 \right)  \left( b_1+3 \right) } \left( b_1+4 \right) }
{ \left( b_1+3 \right)  \left( b_1+2 \right) ^{2}}}O_E(b_0-1,b_1+2)\right]
$$
{\vskip 1cm}
$$ \hat{D}O_D(b_0,b_1) =\sqrt{\left(N-b_0-b_1\right)\left(N-b_0-1\right)}\left[
-{4\over 3}\sqrt {\left(b_1+1\right)\left( b_1-2 \right)\over b_1 \left( b_1-1 \right)}
\,{\frac { \left( b_1-3 \right)  \left( b_1+2 \right) }{{b_1} \left( b_1-1 \right)}}O_D(b_0+1,b_1-2)\right.
$$
$$
\left.
+{2\over 3}\,{\frac { \left( b_1+2 \right) \sqrt {b_1\, \left( b_1-1 \right)  \left( b_1+1 \right)  \left( b_1-2 \right) }\sqrt {2}}
{b_1\, \left( b_1-1 \right) ^{2}}}O_E(b_0+1,b_1-2)
-4\,{\frac { \left( b_1+2 \right) \sqrt {b_1\, \left( b_1-2 \right)}}{{b_1}^{2} 
\left( b_1-1 \right) }}O_F(b_0,b_1)\right]
$$
$$+\sqrt { \left( N-b_0-b_1-1 \right)  \left( N-b_0 \right) }
\left[{\frac {8}{b_1\left( b_1+1 \right)}}\sqrt { \left( b_1+2 \right)  \left( b_1-1 \right)\over  \left( b_1+1 \right) b_1}O_A(b_0+1,b_1-2)
\right.
$$
$$
+{2\sqrt{2}\over 3}\,{\frac {\sqrt{b_1( b_1+2)}(b_1-1)(3b_1+8)}{(b_1+1)(b_1+2)^2 b_1}}O_C(b_0,b_1)
-{32\over 3}{(b_1^2+2b_1-3)\sqrt{b_1(b_1+2)}\over (b_1+2)^2 b_1^2 (b_1+1)}O_B(b_0,b_1)
$$
$$
-{4\over 3}\, \sqrt {{\frac {b_1\, \left( b_1+3 \right) }{ \left( b_1+2 \right)  \left( b_1+1 \right) }}}
{\left( b_1-1 \right)  \left( b_1+4 \right) \over\left(b_1+2\right)\left(b_1+1\right)}O_D(b_0-1,b_1+2)
$$
$$
\left.
+{2\over 3}\,  \sqrt {{\frac {b_1\, \left( b_1+3 \right) }{ \left( b_1+2 \right)  \left( b_1+1 \right) }}}\sqrt {2}
{\left( b_1-1 \right)\over b_1+2 }O_E(b_0-1,b_1+2)\right]
+  \left( N-b_0 \right) \left[
{4\over 3}{\frac {(b_1-1)(b_1^3+b_1^2+16)}{(b_1+1)b_1^2(b_1+2)}}
O_D(b_0,b_1)\right.
$$
$$
+{8\over 3}\, \sqrt {{\frac {b_1-1}{b_1+1}}}{ {b_1}^{2}-4 \over b_1^2 \left( b_1+1 \right)}O_B(b_0+1,b_1-2)
-{4\over 3}\, \sqrt {2}\sqrt {{\frac {b_1-1}{b_1+1}}} {b_1+2 \over b_1^2}O_C(b_0+1,b_1-2)
$$
$$
\left.
-{2\sqrt {2}\over 3}\,{\frac {(b_1^2+2b_1-4)(b_1-1)}{(b_1+1)b_1^2}}O_E(b_0,b_1)
+4\, \left( b_1-1 \right) \sqrt {{\frac {b_1+3}{b_1+1}}}{1\over \left( b_1+2 \right)\left( b_1+1 \right)}
O_F(b_0-1,b_1+2)\right]
$$
$$
+\left( N-b_0-b_1 \right) \left[
{4\over 3}\,{\frac { \left( b_1-2 \right)  \left( b_1+2 \right) ^{2} }{ \left( b_1-1 \right) {b_1}^{2}}}O_D(b_0,b_1)
-{2\sqrt {2}\over 3}\,{\frac { \left( {b_1}^{2}-4 \right) }{{b_1}^{2}}}O_E(b_0,b_1)\right.
$$
$$
\left.
-{8\over 3}\,\sqrt{\left(b_1+1\right)\over \left(b_1-1\right)}
{\frac {\left(b_1^2-4\right)}{{b_1}^{2} \left( b_1-1 \right)}}O_B(b_0+1,b_1-2)
-{2\sqrt {2}\over 3}\,\sqrt{\left(b_1+1\right)\over\left(b_1-1\right)}{\frac{\left(b_1^2-4\right)}{{b_1}^{2} \left( b_1-1 \right)}}
O_C(b_0+1,b_1-2)\right]
$$
{\vskip 1cm}
$$ \hat{D}O_E(b_0,b_1) =\sqrt { \left( N-b_0-b_1 \right)  \left( N-b_0-1 \right) }\left[
{2\sqrt{2}\over 3}\sqrt {\left( b_1+1 \right)  \left( b_1-2 \right) \over b_1\, \left( b_1-1 \right)}
\,{\frac { \left( b_1-3 \right)}
{b_1}}O_D(b_0+1,b_1-2)\right.
$$
$$
\left.
-{2\over 3}\,{\frac {\sqrt {b_1\, \left( b_1-1 \right)  \left( b_1+1 \right)  \left( b_1-2 \right)}}{b_1\, \left( b_1-1 \right) }}
O_E(b_0+1,b_1-2)
+2\,{\frac {\sqrt {2}\sqrt {b_1\, \left( b_1-2 \right) }}{{b_1}^{2}}}
O_F(b_0,b_1)\right]
$$
$$
+\sqrt { \left( N-b_0-b_1-1 \right)  \left( N-b_0 \right) }\left[
2\,{\frac {\sqrt {2}\sqrt { \left( b_1+2 \right)  \left( b_1-1 \right)  
\left( b_1+1 \right) b_1}}{{b_1}^{2} \left( b_1+1 \right) ^{2}}}
O_A(b_0+1,b_1-2)\right.
$$
$$
-{4\over 3}{\sqrt{b_1(b_1+2)}\over (b_1+2)(b_1+1)b_1}O_C(b_0,b_1)
-{2\sqrt {2}\over 3}\sqrt {b_1(b_1+2)}{(b_1+3)(3b_1-2) \over (b_1+1)b_1^2(b_1+2)}O_B(b_0,b_1)
$$
$$
+{2\over 3}\sqrt{{\frac {b_1(b_1+3)}{(b_1+2)(b_1+1)}}}\sqrt {2}
{b_1+4\over b_1+1}O_D(b_0-1,b_1+2)
\left.
-{2\over 3}\,\sqrt {{\frac {b_1\, \left( b_1+3 \right) }{ \left( b_1+1 \right)  \left( b_1+2 \right) }}}
O_E(b_0-1,b_1+2)\right]
$$
$$+\left( N-b_0 \right)\left[
-{2\sqrt {2}\over 3}{\frac {(b_1^2+2b_1-4)(b_1-1)}{(b_1+1)b_1^2}}O_D(b_0,b_1)
\right.
$$
$$
+{2\over 3}\,{\frac {(b_1+2)(b_1^2+2b_1-1)}{(b_1+1)b_1^2}}O_E(b_0,b_1)
-2\,\sqrt {{\frac {b_1+3}{b_1+1}}}\sqrt {2} {1\over b_1+1}O_F(b_0-1,b_1+2)
$$
$$
\left. +{2\over 3}\sqrt {2}\sqrt {{\frac {b_1-1}{b_1+1}}} {b_1^2-4 \over (b_1+1)b_1^2}O_B(b_0+1,b_1-2)
-{2\over 3}\sqrt{{\frac {b_1-1}{b_1+1}}} {b_1+2 \over b_1^2}O_C(b_0+1,b_1-2)\right]
$$
$$
+\left( N-b_0-b_1 \right)\left[{2\over 3}\,{\frac {   \left( {b_1}^{2}-3\,b_1+2 \right) }{{b_1}^{2}}}O_E(b_0,b_1)
-{2\sqrt {2}\over 3}\,{\frac{\left( {b_1}^{2}-4 \right) }{{b_1}^{2}}}O_D(b_0,b_1)\right.
$$
$$
\left.
+{4\sqrt {2}\over 3}\,{\frac { \left( b_1-2 \right) \sqrt { \left( b_1+1 \right)  \left( b_1-1 \right) }}{{b_1}^{2} 
\left( b_1-1 \right) }}O_B(b_0+1,b_1-2)
+{2\over 3}\,{\frac {\left(b_1-2\right)\sqrt{\left(b_1+1\right)\left(b_1-1\right)}}{{b_1}^{2}\left(b_1-1\right)}}
O_C(b_0+1,b_1-2)\right]
$$
{\vskip 1cm}
$$ \hat{D}O_F(b_0,b_1) =\sqrt { \left( N-b_0-b_1 \right)  \left( N-b_0-1 \right) }\left[
8\,{\frac {\sqrt {b_1\, \left( b_1-1 \right)  \left( b_1+1 \right)  \left( b_1-2 \right) }
}{{b_1}^{2} \left( b_1-1 \right) ^{2}}}O_B(b_0+1,b_1-2)\right.
$$
$$
+2\,{\frac {\sqrt {2}
\sqrt {b_1\, \left( b_1-1 \right)  \left( b_1+1 \right)  \left( b_1-2 \right) }}{{b_1}^{2} \left( b_1-1 \right) ^{2}}}O_C(b_0+1,b_1-2)
-4\, \left( b_1+2 \right) \sqrt {{\frac {b_1-2}{b_1}}}
{1\over b_1 \left( b_1-1 \right)}O_D(b_0,b_1)
$$
$$
\left.
+2\,\sqrt {{\frac {b_1-2}{b_1}}}\sqrt {2}{1\over b_1} O_E(b_0,b_1)
\right]
+4\,{\frac { \left( b_1-3 \right) \sqrt { \left( b_1+1 \right)  \left( b_1-1 \right) } \left( N-b_0-1 \right) }{b_1\, \left( b_1-1 \right) ^{2}}}
O_D(b_0+1,b_1-2)
$$
$$
-2\,{\frac {\sqrt { \left( b_1+1 \right)  \left( b_1-1 \right) }\sqrt {2} \left( N-b_0-1 \right) }{ \left( b_1-1 \right) ^{2}}}
O_E(b_0+1,b_1-2)
+12\,{\frac {N-b_0-1}{b_1\, \left( b_1-1 \right) }}O_F(b_0,b_1)
$$
}

\subsection{Four Impurities}

{\footnotesize

$$
\hat{D}O_A(b_0,b_1)=\sqrt{ (N-b_0-b_1-3)(N-b_0+1)}\left[
6\sqrt{b_1+5\over b_1+3}{b_1\over (b_1+4)(b_1+2)}
O_B(b_0,b_1)\right.
$$
$$
-2\sqrt{3}\sqrt{b_1+5\over b_1+3} {1\over b_1+2}O_C(b_0,b_1)
+12{\sqrt{(b_1+1)(b_1+5)}\over (b_1+3)(b_1+2)(b_1+4)} O_D(b_0-1,b_1+2)
$$  
$$
\left. + 4\sqrt {3}{\sqrt{(b_1+5)(b_1+1)}\over (b_1+3)(b_1+4)(b_1+2)}
O_E(b_0-1,b_1+2)\right]
+(N-b_0-b_1-3)\left[
{\frac {24}{(b_1+2)(b_1+4)}}O_A(b_0,b_1)\right.
$$
$$
\left.
-6\sqrt{b_1+1\over b_1+3}{\frac{b_1+6}{(b_1+2)(b_1+4)}}O_B(b_0-1,b_1+2)
+ 2\sqrt {3}\sqrt{b_1+1\over b_1+3}{\frac {1}{b_1+4}}O_C(b_0-1,b_1+2)\right]
$$
{\vskip 1cm}
$$
\hat{D}O_B(b_0,b_1)=\sqrt { \left( N-b_0-b_1-2 \right)\left( N-b_0 \right) }\left[
{\sqrt{3}\over 2}\sqrt{\left(b_1+3\right)(b_1-1)}{b_1+4\over b_1(b_1+1)}O_C(b_0+1,b_1-2)
\right.
$$
$$
-{3\over 2} {\sqrt{(b_1+3)(b_1-1)}\over (b_1+1)(b_1+2)}{(b_1+4)(b_1-2)\over b_1}O_B(b_0+1,b_1-2)
+3 {(b_1+4)(b_1-1)(b_1-6)\over b_1(b_1+3)(b_1+2)^2}\sqrt{b_1+3\over b_1+1}O_D(b_0,b_1)
$$
$$
-\sqrt {3} {(b_1+4)(3b_1-2)\over b_1(b_1+2)^2} \sqrt{b_1+3\over b_1+1}O_E(b_0,b_1)
%
%
\left. +
{2b_1(b_1+4)\over (b_1+1)(b_1+3)(b_1+2)^2}\left(9 O_G(b_0-1,b_1+2)+\sqrt {3}O_H(b_0-1,b_1+2)\right)\right]
$$
$$
+\sqrt {(N-b_0-b_1-3)(N-b_0+1)}\left[
-{3\over 2}\sqrt{(b_1+5)(b_1+1)}{b_1(b_1+6)\over (b_1+3)(b_1+2)(b_1+4)} O_B(b_0-1,b_1+2)\right.
$$
$$
+{\sqrt{3}\over 2}\sqrt{(b_1+5)(b_1+1)}{b_1\over (b_1+3)(b_1+4)}
O_C(b_0-1,b_1+2)\left.
+6\sqrt{(b_1+5)(b_1+3)}{\frac{b_1}{(b_1+2)(b_1+3)(b_1+4)}}O_A(b_0,b_1)\right]
$$
$$
+(N-b_0-b_1-2)\left[
-6\sqrt {{\frac {b_1-1}{b_1+1}}}{b_1+4\over b_1(b_1+2)}O_A(b_0+1,b_1-2)
+{3\over 2}{\frac {(b_1^3+7b_1^2+22b_1-24)(b_1+4)}{b_1(b_1+2)^2(b_1+3)}}O_B(b_0,b_1)\right.
$$
$$
-{\sqrt {3}\over 2}{\frac{(b_1^2+3b_1-6)(b_1+4)}{(b_1+2)^2(b_1+3)}}O_C(b_0,b_1)
-6 \sqrt {{\frac {b_1+3}{b_1+1}}}{b_1(b_1+5)(b_1+4)\over (b_1+2)^2 (b_1+3)^2}
O_D(b_0-1,b_1+2)
$$
$$
\left.
+\sqrt {12}\sqrt{(b_1+3)(b_1+1)}{b_1(b_1+4)\over (b_1+3)^2 (b_1+2)^2}
O_E(b_0-1,b_1+2)\right]+(N-b_0+1)\left[
{3\over 2}{\frac{b_1^2(b_1+5)}{(b_1+2)(b_1+3)(b_1+4)}}O_B(b_0,b_1)\right.
$$
$$
-{\sqrt{3}\over 2}{\frac{(b_1+5)b_1}{(b_1+2)(b_1+3)}}
O_C(b_0,b_1)
+3 \sqrt{(b_1+3)(b_1+1)}{b_1(b_1+5)\over (b_1+3)^2(b_1+4)(b_1+2)}
O_D(b_0-1,b_1+2)
$$
$$
\left. +\sqrt {3}\sqrt{{(b_1+3)(b_1+1)}}{b_1(b_1+5)\over (b_1+3)^2 (b_1+2)(b_1+4)}O_E(b_0-1,b_1+2)\right]
$$
{\vskip 1cm}
$$
\hat{D}O_C(b_0,b_1)=\sqrt{(N-b_0-b_1-2)( N-b_0)}\left[
{\sqrt{3}\over 2}
\sqrt {(b_1-1)( b_1+3)}{b_1-2\over (b_1+1)(b_1+2)}O_B(b_0+1,b_1-2)
\right.
$$
$$
-{1\over 2}\sqrt{(b_1+3)(b_1-1)}{1 \over b_1+1}O_C(b_0+1,b_1-2)
+{1\over\sqrt{3}}\sqrt{b_1+3\over b_1+1} {(b_1-1)(5b_1+18) \over (b_1+3)(b_1+2)^2}O_D(b_0,b_1)
$$
$$
+\sqrt{b_1+3\over b_1+1}{3b_1-2\over (b_1+2)^2}O_E(b_0,b_1)
+ 2\sqrt{3}{\frac{b_1(b_1+4)}{(b_1+1)(b_1+2)^2(b_1+3)}}
O_G(b_0-1,b_1+2)
$$
$$
-{4\sqrt{6}\over 3}\sqrt{b_1+3\over b_1+1}{1\over b_1+2}
O_F(b_0,b_1)
%
+2\left. {\frac{(3b_1^2+12b_1+8)}{(b_1+3)(b_1+1)(b_1+2)^2}}O_H(b_0-1,b_1+2)\right]
$$
$$
+\sqrt{(N-b_0-b_1-3)(N-b_0+1)}\left[
{\sqrt{3}\over 2}\sqrt{(b_1+5)(b_1+1)}
{\frac {b_1+6}{(b_1+3)(b_1+2)}}O_B(b_0-1,b_1+2)\right.
$$
$$
-{1\over 2}{\frac{\sqrt{(b_1+5)(b_1+1)}}{(b_1+3)}}O_C(b_0-1,b_1+2)
%
\left. 
-2\sqrt {3}{\frac{\sqrt{(b_1+5)(b_1+3)}}{( b_1+3)(b_1+2)}}O_A(b_0,b_1)\right]
+(N-b_0-b_1-2)\times
$$
$$
\times\left[
\sqrt{{\frac{b_1-1}{b_1+1}}}{2\sqrt {3}\over b_1+2}O_A(b_0+1,b_1-2)\right.
%
-{\sqrt{3}\over 2}{\frac{(b_1^2+3b_1-6)(b_1+4)}{(b_1+2)^2(b_1+3)}}O_B(b_0,b_1)
+{\frac{b_1^3+3b_1^2+10b_1+32}{2(b_1+2)^2(b_1+3)}}O_C(b_0,b_1)
$$
$$
-{2\over \sqrt{3}}
\sqrt {{\frac {b_1+3}{b_1+1}}}{b_1(b_1+5)(b_1+4)\over (b_1+2)^2(b_1+3)^2}
O_D(b_0-1,b_1+2)
-2\sqrt{(b_1+3)(b_1+1)}{(b_1+4)^2\over (b_1+2)^2(b_1+3)^2}O_E(b_0-1,b_1+2)
$$
$$
+\left. {4\sqrt{2}\over\sqrt{3}}{\sqrt {(b_1+3)(b_1+1)}\over (b_1+2)(b_1+3)}
O_F(b_0-1,b_1+2)\right]+(N-b_0+1)\left[
-{\sqrt {3}\over 2}{\frac{b_1(b_1+5)}{(b_1+2)(b_1+3)}}O_B(b_0,b_1)\right.
$$
$$
+{1\over 2}{\frac{(b_1+5)(b_1+4)}{(b_1+2)(b_1+3)}}O_C(b_0,b_1)
-\sqrt{3}\sqrt {(b_1+3)(b_1+1)}{b_1+5\over (b_1+3)^2 (b_1+2)}
O_D(b_0-1,b_1+2)
$$
$$
\left. -\sqrt{(b_1+3)(b_1+1)}{(b_1+5)\over (b_1+2)(b_1+3)^2}
O_E(b_0-1,b_1+2)\right]
$$
{\vskip 1cm}
$$
\hat{D}O_D(b_0,b_1)=
\sqrt {(N-b_0-b_1-1)(N-b_0-1)}\left[
-2{(b_1^2-9)(b_1^2-4)\over b_1^2(b_1^2 -1)}
O_D(b_0+1,b_1-2)
\right.
$$
$$
\left.
+{(b_1^2-4)(b_1+3)\over b_1^2 (b_1+1)^2}\left[
{2(b_1+1)^2\over\sqrt {3}(b_1-1)}O_E(b_0+1,b_1-2)
-\sqrt{b_1+1\over b_1-1}\left(6O_G(b_0,b_1)+{2\over\sqrt{3}}O_H(b_0,b_1) \right)\right]\right]
$$
$$
+3\sqrt{b_1-1\over b_1+1} {{b_1}^{2}+b_1-6\over b_1(b_1+1)(b_1+2)}
O_G(b_0,b_1)
-\sqrt {3}\sqrt{b_1-1\over b_1+1}
{b_1+3\over b_1(b_1+1)}
O_H(b_0,b_1)
$$
$$
+ \left. 12{\frac {\sqrt { ( b_1+3) ( b_1-1)}}{b_1 ( b_1+1) ( b_1+2)}}O_I(b_0-1,b_1+2)\right]
+\sqrt {(N-b_0-b_1-2)(N-b_0)}\times
$$
$$
\times\left[
-3\sqrt{(b_1+3)(b_1+1)}{\frac {(b_1+4)(b_1-1)}{(b_1+1)^2 b_1 (b_1+2)}}O_B(b_0,b_1)
+\sqrt {3}\sqrt{(b_1+3)(b_1+1)}{\frac {b_1-1}{(b_1+1)^2 (b_1+2)}}O_C(b_0,b_1)\right.
$$
$$
+12 {\frac {\sqrt{(b_1+3)(b_1-1)}}{b_1(b_1+1)(b_1+2)}}O_A(b_0+1,b_1-2)
-2{b_1(b_1-1)(b_1+4)(b_1+5)\over (b_1+1)(b_1+3)(b_1+2)^2}
O_D(b_0-1,b_1+2)
$$
$$
+6\sqrt {b_1+1\over b_1+3}
{\frac {b_1(b_1^2+3b_1-4)}{(b_1+2)^2(b_1+1)^2}}
O_B(b_0,b_1)
+{2\over \sqrt{3}}\sqrt{b_1+1\over b_1+3}
{\frac {b_1(b_1^2+3b_1-4)}{(b_1+2)^2 (b_1+1)^2}}
O_C(b_0,b_1)
$$
$$
\left.+
{2\over \sqrt{3}}
{b_1(b_1+4)(b_1-1)\over (b_1+3)(b_1+2)^2}
O_E(b_0-1,b_1+2)\right]
+(N-b_0-b_1-1)\left[
-6\sqrt {{\frac {b_1-1}{b_1+1}}}{b_1^3+3b_1^2-4b_1-12\over b_1^2(b_1^2-1)}
O_B(b_0+1,b_1-2)\right.
$$
$$
-{2\over\sqrt{3}}\sqrt {{\frac {b_1-1}{b_1+1}}}{b_1^3+3b_1^2-4b_1-12\over b_1^2(b_1^2-1)}
O_C(b_0+1,b_1-2)
+2{\frac{(b_1-2)(b_1+3)^2(b_1+2)}{b_1^2(b_1+1)^2}}
O_D(b_0,b_1)
$$
$$
\left.
-{2\over\sqrt{3}}{\frac{(b_1-2)(b_1+3)(b_1-1)(b_1+2)}{b_1^2(b_1+1)^2}}
O_E(b_0,b_1)\right]+(N-b_0)\left[
3\sqrt{{\frac {b_1-1}{b_1+1}}}{b_1^2+b_1-6\over b_1(b_1+1)(b_1+2)}O_B(b_0+1,b_1-2)\right.
$$
$$
-\sqrt{3}\sqrt{{\frac {b_1-1}{b_1+1}}}{b_1+3\over b_1(b_1+1)}O_C(b_0+1,b_1-2)
+6{\frac{(b_1+3)(b_1-1)}{b_1(b_1+2)(b_1+1)^2}}O_D(b_0,b_1)
$$
$$
\left. +2\sqrt {3}{\frac{(b_1+3)(b_1-1)}{(b_1+2)b_1(b_1+1)^2}}O_E(b_0,b_1)\right]
+(N-b_0-b_1-1)\left[
6{\frac{(b_1+3)(b_1-1)}{(b_1+1)^2 b_1 (b_1+2)}}O_D(b_0,b_1)\right.
$$
$$
+2\sqrt{3}{\frac{(b_1-1)(b_1+3)}{(b_1+1)^2b_1( b_1+2)}}O_E(b_0,b_1)
-3\sqrt{{\frac{b_1+3}{b_1+1}}}{b_1^2+3b_1-4\over b_1(b_1+1)(b_1+2)}O_G(b_0-1,b_1+2)
$$
$$
\left.
+\sqrt{3}\sqrt{{\frac {b_1+3}{b_1+1}}}{(b_1-1)\over (b_1+1)(b_1+2)}O_H(b_0-1,b_1+2)\right]
+(N-b_0)\left[
2{\frac {(b_1+4)(b_1-1)^2b_1}{(b_1+1)^2(b_1+2)^2}}
O_D(b_0,b_1)\right. 
$$
$$
-{2\over\sqrt{3}}{\frac{(b_1-1)(b_1+3)b_1(b_1+4)}{(b_1+1)^2(b_1+2)^2}}
O_E(b_0,b_1)
+6\sqrt{{\frac{b_1+3}{b_1+1}}}{(b_1^2+3b_1-4)b_1 \over(b_1+1)(b_1+2)^2(b_1+3)}
O_G(b_0-1,b_1+2)
$$
$$
\left. +{2\over \sqrt{3}}\sqrt{{\frac {b_1+3}{b_1+1}}}{(b_1^2+3b_1-4)b_1 \over (b_1+1)(b_1+2)^2(b_1+3)}
O_H(b_0-1,b_1+2)  \right]
$$
{\vskip 1cm}
$$
\hat{D}O_E(b_0,b_1)=\sqrt{(N-b_0-b_1-1)( N-b_0-1)}\left[
{2\over\sqrt{3}}
{b_1^3-3b_1^2-4b_1+12\over (b_1+1)b_1^2}
O_D(b_0+1,b_1-2)\right.
$$
$$
-2{(b_1^2-4)\over b_1^2}O_E(b_0+1,b_1-2)
+{2\sqrt{6}\over 3}{b_1+2\over b_1}
O_F(b_0+1,b_1-2)
+\sqrt{3}\sqrt{b_1-1\over b_1+1}{(b_1-2)(3b_1+8)\over b_1^2(b_1+2)}O_G(b_0,b_1)
$$
$$
\left.
-\sqrt{b_1-1\over b_1+1}{\frac{3b_1+8}{b_1^2}}O_H(b_0,b_1)
+4\sqrt{3}{\frac{\sqrt{(b_1+3)(b_1-1)}}{b_1(b_1+1)(b_1+2)}}O_I(b_0-1,b_1+2)\right]
$$
$$
+\sqrt {(N-b_0-b_1-2)(N-b_0)}
\left[-\sqrt {3}\sqrt{(b_1+1)(b_1+3)}{\frac{(b_1+4)(b_1-1)}{b_1(b_1+1)^2(b_1+2)}}O_B(b_0,b_1)\right.
$$
$$
+\sqrt{(b_1+1)(b_1+3)}{\frac {b_1-1}{(b_1+2)(b_1+1)^2}}O_C(b_0,b_1)
+4\sqrt{3}{\frac {\sqrt{(b_1-1)(b_1+3)}}{b_1(b_1+1)(b_1+2)}}
O_A(b_0+1,b_1-2)
$$
$$
-2\sqrt{3}\sqrt{(b_1+3)(b_1+1)}
{\frac {b_1(b_1+4)}{(b_1+1)^2(b_1+2)^2}}
O_B(b_0,b_1)
$$
$$
+2\sqrt{(b_1+1)(b_1+3)}{\frac {b_1^2}{(b_1+2)^2(b_1+1)^2}}O_C(b_0,b_1)
+{2\over\sqrt{3}}
{\frac{b_1 (b_1^2+9b_1+20)}{(b_1+1)(b_1+2)^2}} O_D(b_0-1,b_1+2)
$$
$$
\left.
-2{\frac {b_1(b_1+4)}{(b_1+2)^2}}O_E(b_0-1,b_1+2)
+{2\sqrt {2}\over\sqrt{3}}
{\frac {b_1}{(b_1+2)}}O_F(b_0-1,b_1+2)\right]+(N-b_0-b_1-1)\times
$$
$$
\times\left[
2\sqrt {3}\sqrt {{\frac {b_1-1}{b_1+1}}}{b_1^2-4\over b_1^2(b_1+1)}
O_B(b_0+1,b_1-2)
-2\sqrt {{\frac {b_1-1}{b_1+1}}}{(b_1+2)^2 \over b_1^2(b_1+1)}O_C(b_0+1,b_1-2)\right.
$$
$$
-{2\over\sqrt{3}}{\frac{(b_1-2)(b_1+3)(b_1-1)(b_1+2)}{b_1^2(b_1+1)^2}}
O_D(b_0,b_1)
+2{\frac{(b_1^4+2b_1^3+b_1^2-4)}{b_1^2 (b_1+1)^2}}
O_E(b_0,b_1)
$$
$$
\left.
-{2\sqrt{6}\over 3}{\frac{b_1^2+b_1-2}{b_1(b_1+1)}}
O_F(b_0,b_1)\right]+(N-b_0)\left[
\sqrt{3}\sqrt{{\frac {b_1-1}{b_1+1}}}{b_1^2+b_1-6\over (b_1+2)b_1(b_1+1)}O_B(b_0+1,b_1-2)\right.
$$
$$
-\sqrt{{\frac{b_1-1}{b_1+1}}}{b_1+3\over b_1(b_1+1)}O_C(b_0+1,b_1-2)
+2\sqrt{3}{\frac{(b_1+3)(b_1-1)}{b_1(b_1+2)(b_1+1)^2}}O_D(b_0,b_1)
$$
$$
\left.
+2{\frac{(b_1+3)(b_1-1)}{b_1(b_1+2)(b_1+1)^2}}O_E(b_0,b_1)\right]
+(N-b_0-b_1-1)\left[
2\sqrt{3}{\frac{(b_1-1)(b_1+3)}{(b_1+1)^{2}(b_1+2)b_1}}
O_D(b_0,b_1)
\right.
$$
$$
+2{\frac{(b_1+3)(b_1-1)}{(b_1+1)^2b_1(b_1+2)}}O_E(b_0,b_1)
-\sqrt{3}\sqrt{{\frac {b_1+3}{b_1+1}}}{b_1^2+3b_1-4\over b_1(b_1+1)(b_1+2)}
O_G(b_0-1,b_1+2)
$$
$$
\left.
+\sqrt{{\frac{b_1+3}{b_1+1}}}{b_1-1\over (b_1+1)(b_1+2)}
O_H(b_0-1,b_1+2)\right]+(N-b_0)\left[
-{2\over\sqrt{3}}{\frac{(b_1-1)(b_1+3)b_1(b_1+4)}{(b_1+1)^2(b_1+2)^2}}
O_D(b_0,b_1)\right.
$$
$$
+2{\frac{(b_1+3)(b_1^2+3b_1+4)b_1}{(b_1+1)^2(b_1+2)^2}}O_E(b_0,b_1)
-{2\sqrt{6}\over 3}{\frac{(b_1+3)b_1}{(b_1+2)(b_1+1)}}O_F(b_0,b_1)
$$
$$
\left.
-2\sqrt{3}\sqrt{{\frac{b_1+3}{b_1+1}}} {(b_1+4)b_1 \over (b_1+1)(b_1+2)^2}
O_G(b_0-1,b_1+2)
+2\sqrt{{\frac {b_1+3}{b_1+1}}}{b_1^2\over (b_1+2)^2(b_1+1)}
O_H(b_0-1,b_1+2)\right]
$$
{\vskip 1cm}
$$
\hat{D}O_F(b_0,b_1)=\sqrt{(N-b_0-b_1-1)(N-b_0-1)}\left[
{2\sqrt{6}\over 3}{b_1-2\over b_1}O_E(b_0+1,b_1-2)\right.
-2O_F(b_0+1,b_1-2)
$$  
$$
\left.
+{4\sqrt {6}\over 3}{\frac{\sqrt{(b_1-1)(b_1+1)}}{b_1(b_1+1)}}
O_H(b_0,b_1)\right]
%
+\sqrt {(N-b_0-b_1-2)(N-b_0)}\Big[
-2O_F(b_0-1,b_1+2)
$$
$$
\left.
{2\sqrt{2}\over\sqrt{3}} {\frac {b_1+4}{b_1+2}}O_E(b_0-1,b_1+2)
-{4\sqrt{6}\over 3}\sqrt{b_1+3 \over b_1+1}{1\over b_1+2} O_C(b_0,b_1)\right]+(N-b_0-b_1-1)\times
$$
$$
\times \left[
{4\sqrt{6}\over 3}\sqrt {{\frac {b_1-1}{b_1+1}}} {1\over b_1} O_C(b_0+1,b_1-2)
%
-{2\sqrt{6}\over 3}{\frac{b_1^2+b_1-2}{b_1(b_1+1)}}O_E(b_0,b_1)
+2{\frac{b_1-1}{b_1+1}}O_F(b_0,b_1)\right]
$$
$$
+(N-b_0)\left[
-{2\sqrt{6}\over 3}{\frac{(b_1+3)b_1}{(b_1+2)(b_1+1)}}O_E(b_0,b_1)
+2{\frac{(b_1+3)}{b_1+1}}O_F(b_0,b_1)\right.
%
\left. -{4\sqrt{6}\over 3}\sqrt{{\frac {b_1+3}{b_1+1}}}{1\over b_1+2}
O_H(b_0-1,b_1+2)\right]
$$
{\vskip 1cm}
$$
\hat{D}O_G(b_0,b_1)=\sqrt{(N-b_0-b_1)(N-b_0-2)}\left[
-{3\over 2}{\frac{\sqrt{(b_1+1)(b_1-3)}(b_1^2-2b_1-8)}{b_1(b_1-2)( b_1-1)}}
O_G(b_0+1,b_1-2)\right.
$$
$$
+{\sqrt{3}\over 2}{\frac{\sqrt{(b_1+1)(b_1-3)}(b_1+2)}{(b_1-2)(b_1-1)}}
O_H(b_0+1,b_1-2)
\left.
-6\sqrt {b_1-3\over b_1-1}{\frac {(b_1+2)}{b_1(b_1-2)}}
O_I(b_0,b_1)
\right]
$$
$$
+\sqrt{(N-b_0-b_1-1)(N-b_0-1)}\left[
18{\frac{({b_1}^2-4)}{b_1^2(b_1^2-1)}}
O_B(b_0+1,b_1-2)\right.
$$
$$
+ 2\sqrt {3}{\frac{(b_1^2-4)}{b_1^2(b_1^2-1)}}O_C(b_0+1,b_1-2)
-3\sqrt {b_1-1\over b_1+1}{\frac {(b_1-2)(b_1+8)(b_1+3)}{b_1^2(b_1+2)(b_1-1)}}O_D(b_0,b_1)
$$
$$
+\sqrt {3}\sqrt {{\frac {b_1-1}{b_1+1}}} {(b_1-2)(3b_1+8)\over b_1^2(b_1+2)}
O_E(b_0,b_1)
-{3\over 2}\sqrt {(b_1+3)(b_1-1)}
{\frac{(b_1^2+2b_1-8)}{b_1(b_1+1)(b_1+2)}}O_G(b_0-1,b_1+2)
$$
$$
\left.
+{\sqrt{3}\over 2}\sqrt{(b_1+3)(b_1-1)}{\frac{(b_1-2)}{(b_1+1)(b_1+2)}} O_H(b_0-1,b_1+2)\right]
+(N-b_0-b_1)\left[
-{\sqrt{3}\over 2}{\frac{b_1^2-b_1-6}{b_1(b_1-1)}}O_H(b_0,b_1)
\right.
$$
$$
-\sqrt{3}\sqrt{(b_1+1)(b_1-1)}{\frac{b_1^2-b_1-6}{(b_1-2)(b_1-1)^2b_1}}
O_E(b_0+1,b_1-2)
+{3\over 2}{\frac{(b_1-3)(b_1+2)^2}{b_1(b_1-2)(b_1-1)}}
O_G(b_0,b_1)
$$
$$
\left.
-3\sqrt{b_1+1\over b_1-1}{\frac{b_1^2-b_1-6}{(b_1-2)(b_1-1)b_1}}O_D(b_0+1,b_1-2)\right]
+(N-b_0-1)\left[
-{\sqrt{3}\over 2}{\frac{(b_1-2)(b_1^2+b_1-8)}{b_1^2(b_1-1)}}O_H(b_0,b_1)\right.
$$
$$
$$
$$
+{3\over 2}{\frac{(b_1-2)(b_1^3-b_1^2+6b_1+48)}{b_1^2(b_1-1)(b_1+2)}}O_G(b_0,b_1)
+6\sqrt{{\frac {b_1-1}{b_1+1}}}{(b_1-2)(b_1-3)(b_1+2)\over b_1^2(b_1-1)^2}
O_D(b_0+1,b_1-2)
$$
$$
-2\sqrt{3}\sqrt{(b_1+1)(b_1-1)}{\frac{b_1^2-4}{(b_1-1)^2 b_1^2}}
O_E(b_0+1,b_1-2)
\left.
+6\sqrt{{\frac{b_1+3}{b_1+1}}}{( b_1-2)\over b_1(b_1+2)} O_I(b_0-1,b_1+2)\right]
$$
{\vskip 1cm}
$$
\hat{D}O_H(b_0,b_1)=\sqrt{(N-b_0-b_1)(N-b_0-2)}\left[
{\sqrt{3}\over 2}
{\sqrt{(b_1+1)(b_1-3)}(b_1-4)\over (b_1-1)b_1}
O_G(b_0+1,b_1-2)\right.
$$
$$
\left.
-{1\over 2}\,{\frac{\sqrt{(b_1+1)(b_1-3)}}{b_1-1}}
O_H(b_0+1,b_1-2)
+2\sqrt {3}{\frac{\sqrt{(b_1-1)(b_1-3)}}{b_1( b_1-1)}}
O_I(b_0,b_1)\right]
$$
$$
+\sqrt{(N-b_0-b_1-1)(N-b_0-1)}\left[2\sqrt {3}
{\frac{b_1^2-4}{b_1^2(b_1^2-1)}}O_B(b_0+1,b_1-2)\right.
$$
$$
+2{\frac{3b_1^2-4}{b_1^2(b_1^2-1)}}O_C(b_0+1,b_1-2)
-{2\over\sqrt{3}}\sqrt {b_1-1\over b_1+1}
{\frac{b_1^3+3b_1^2-4b_1-12}{(b_1^2-1)b_1^2}}
O_D(b_0,b_1)
$$
$$
-2\sqrt{b_1-1\over b_1+1}{(b_1+2)^2\over b_1^2 (b_1+1)}
O_E(b_0,b_1)
+{4\sqrt {6}\over 3} \sqrt{b_1-1\over b_1+1}{1\over b_1}
O_F(b_0,b_1)
$$
$$
+\sqrt{b_1-1\over b_1+1 }{b_1+3\over b_1(b_1+1)}
\left(
-\sqrt {3}O_D(b_0,b_1)
-O_E(b_0,b_1)\right)
$$
$$
+\left.
{\sqrt {3}\over 2}\sqrt{(b_1+3)(b_1-1)}{\frac {b_1+4}{b_1(b_1+1)}}O_G(b_0-1,b_1+2)
-{1\over 2}\sqrt{(b_1+3)(b_1-1)}{\frac{1}{b_1+1}}O_H(b_0-1,b_1+2)\right]
$$
$$
+(N-b_0-b_1)\left[\sqrt {3}\sqrt{b_1+1\over b_1-1}{\frac{(b_1-3)}{b_1(b_1-1)}}O_D(b_0+1,b_1-2)
+\sqrt{b_1+1\over b_1-1}{\frac{b_1-3}{b_1(b_1-1)}}O_E(b_0+1,b_1-2)\right.
$$
$$
\left.
-{\sqrt{3}\over 2}{\frac{b_1^2-b_1-6}{b_1(b_1-1)}}
O_G(b_0,b_1)
+{1\over 2}{\frac{b_1^2-5b_1+6}{b_1(b_1-1)}}
O_H(b_0,b_1)\right]+(N-b_0-1)\times
$$
$$
\times\left[ {2\over \sqrt {3}}\sqrt {{\frac {b_1-1}{b_1+1}}}{b_1^3-3b_1^2-4b_1+12\over b_1^2(b_1-1)^2} O_D(b_0+1,b_1-2)
+2{\frac{(b_1^2-4b_1+4)\sqrt{(b_1+1)(b_1-1)}}{b_1^2(b_1-1)^2}} O_E(b_0+1,b_1-2)\right.
$$
$$
-{4\sqrt{6}\over 3}{\frac{\sqrt{(b_1+1)(b_1-1)}}{b_1( b_1-1)}} O_F(b_0+1,b_1-2)
-{\sqrt{3}\over 2}{\frac{(b_1-2)(b_1^2+b_1-8)}{b_1^2(b_1-1)}} O_G(b_0,b_1)
$$
$$
\left. +{\frac{b_1^3+3b_1^2+10b_1-16}{2b_1^2(b_1-1)}} O_H(b_0,b_1)
-2\sqrt {3}\sqrt{{\frac{b_1+3}{b_1+1}}}{1\over b_1} O_I(b_0-1,b_1+2)\right]
$$
{\vskip 1cm}
$$ \hat{D}O_I {(b_0,b_1)}=\sqrt { ( N-b_0-b_1) ( N-b_0-2)}\left[
12\sqrt{(b_1+1)(b_1-3)}{\frac {1}{b_1(b_1-2)(b_1-1)}}
O_D(b_0+1,b_1-2)\right.
$$
$$
+4\sqrt {3}{\frac {\sqrt {(b_1+1)(b_1-3)}}{b_1(b_1-1)(b_1-2)}}O_E(b_0+1,b_1-2)
-6\sqrt{{\frac{b_1-3}{b_1-1}}}{b_1+2\over b_1(b_1-2)}O_G(b_0,b_1)
$$
$$
+\left.
2\sqrt{3}\sqrt{{\frac{b_1-3}{b_1-1}}}{1\over b_1}O_H(b_0,b_1)\right]
+(N-b_0-2)\left[ 6{\frac{\sqrt{(b_1+1)(b_1-1)}(b_1-4)}{b_1(b_1-1)(b_1-2)}}
O_G(b_0+1,b_1-2)\right.
$$
$$
\left. -2\sqrt{3}{\frac{\sqrt{(b_1+1)(b_1-1)}}{(b_1-2)(b_1-1)}}O_H(b_0+1,b_1-2)
+24{\frac {1}{b_1(b_1-2)}}O_I(b_0,b_1)\right]
$$
}

\section{A Discussion on Intertwiners}

When $S^n$ acts on $V^{\otimes n}$ $n>1$ it furnishes a reducible representation. Imagine that this includes the 
irreducible representations $R$ and $T$. Representing the action of $\sigma$ as a matrix $\Gamma(\sigma )$, 
in a suitable basis we can write
$$
\Gamma(\sigma)=\left[
\matrix{\Gamma_R(\sigma) &0 &\cdots\cr 0 &\Gamma_S(\sigma) &\cdots\cr \cdots &\cdots &\cdots}\right]\, .
$$
If we restrict ourselves to an $S_{n-1}$ subgroup of $S_n$, then in general, both $R$ and $S$ will subduce a number of
representations. Assume for the sake of this discussion that $R$ subduces $R_1'$ and $R_2'$ and that
$S$ subduces $S_1'$ and $S_2'$. Then, for $\sigma\in S_{n-1}$ we have
$$
\Gamma(\sigma)=
\left[
\matrix{
\Gamma_{R_1'}(\sigma) &0 &0 &0 &\cdots\cr 
0 &\Gamma_{R_2'}(\sigma) &0 &0 &\cdots\cr 
0 &0 &\Gamma_{S_1'}(\sigma) &0 &\cdots\cr
0 &0 &0 &\Gamma_{S_2'}(\sigma) &\cdots\cr
\cdots &\cdots &\cdots &\cdots &\cdots}\right]\, .
$$
Imagine that $S_1'=R_1'$, that is, one of the irreducible representations subduced by $R$ is
also subduced by $S$. Then, a simple application of the fundamental orthogonality relation gives
$$
\sum_{\sigma\in S_{n-1}}
\left[
\matrix{
\Gamma_{R_1'}(\sigma) &0 &0 &0 &\cdots\cr 
0 &0 &0 &0 &\cdots\cr 
0 &0 &0 &0 &\cdots\cr
0 &0 &0 &0 &\cdots\cr
\cdots &\cdots &\cdots &\cdots &\cdots}\right]_{ij}
\left[
\matrix{
0 &0 &0 &0 &\cdots\cr 
0 &0 &0 &0 &\cdots\cr 
0 &0 &\Gamma_{S_1'}(\sigma) &0 &\cdots\cr
0 &0 &0 &0 &\cdots\cr
\cdots &\cdots &\cdots &\cdots &\cdots}\right]_{ab}
$$
$$
={(n-1)!\over d_{R_1'}}\delta_{R_1'S_1'}
\left[
\matrix{
0 &0 &{\bf 1} &0 &\cdots\cr 
0 &0 &0 &0 &\cdots\cr 
0 &0 &0 &0 &\cdots\cr
0 &0 &0 &0 &\cdots\cr
\cdots &\cdots &\cdots &\cdots &\cdots}\right]_{ib}
\left[
\matrix{
0 &0 &0 &0 &\cdots\cr 
0 &0 &0 &0 &\cdots\cr 
{\bf 1} &0 &0 &0 &\cdots\cr
0 &0 &0 &0 &\cdots\cr
\cdots &\cdots &\cdots &\cdots &\cdots}\right]_{aj}
$$
$$
\equiv {(n-1)!\over d_{R_1'}}\delta_{R_1'S_1'} (I_{R_1'S_1'})_{ib}(I_{S_1'R_1'})_{aj}
$$
where the form of the intertwiners has been spelled out.


\begin{thebibliography}{30}
\parskip-2pt

\bibitem{Corley:2001zk}
  S.~Corley, A.~Jevicki and S.~Ramgoolam,
  ``Exact correlators of giant gravitons from dual N = 4 SYM theory,''
  Adv.\ Theor.\ Math.\ Phys.\  {\bf 5}, 809 (2002)
  [arXiv:hep-th/0111222].

\bibitem{Balasubramanian:2004nb}
  V.~Balasubramanian, D.~Berenstein, B.~Feng and M.~x.~Huang,
  ``D-branes in Yang-Mills theory and emergent gauge symmetry,''
  JHEP {\bf 0503}, 006 (2005)
  [arXiv:hep-th/0411205].

\bibitem{de Mello Koch:2007uu}
  R.~de Mello Koch, J.~Smolic and M.~Smolic,
  ``Giant Gravitons - with Strings Attached (I),'' JHEP {\bf 0706}, 074 (2007),
  arXiv:hep-th/0701066.

\bibitem{de Mello Koch:2007uv}
  R.~de Mello Koch, J.~Smolic and M.~Smolic,
  ``Giant Gravitons - with Strings Attached (II),'' JHEP {\bf 0709} 049 (2007),
  arXiv:hep-th/0701067.

\bibitem{Kimura:2007wy}
  Y.~Kimura and S.~Ramgoolam,
  ``Branes, Anti-Branes and Brauer Algebras in Gauge-Gravity duality,''
  arXiv:0709.2158 [hep-th].

\bibitem{Bekker:2007ea}
  D.~Bekker, R.~de Mello Koch and M.~Stephanou,
  ``Giant Gravitons - with Strings Attached (III),''
  arXiv:0710.5372 [hep-th].

\bibitem{Brown:2007xh}
  T.~W.~Brown, P.~J.~Heslop and S.~Ramgoolam,
  ``Diagonal multi-matrix correlators and BPS operators in N=4 SYM,''
  arXiv:0711.0176 [hep-th].

\bibitem{Bhattacharyya:2008rb}
  R.~Bhattacharyya, S.~Collins and R.~d.~M.~Koch,
  ``Exact Multi-Matrix Correlators,''
  JHEP {\bf 0803}, 044 (2008)
  [arXiv:0801.2061 [hep-th]].

\bibitem{Brown:2008rr}
  T.~W.~Brown, P.~J.~Heslop and S.~Ramgoolam,
  ``Diagonal free field matrix correlators, global symmetries and giant
  gravitons,''
  arXiv:0806.1911 [hep-th].

\bibitem{Kimura:2008wy}
 Y.~Kimura and S.~Ramgoolam,
  ``Enhanced symmetries of gauge theory and resolving the spectrum of local
  operators,''
  Phys.\ Rev.\  D {\bf 78}, 126003 (2008)
  [arXiv:0807.3696 [hep-th]].

\bibitem{Kimura:2009wy}
 Y.~Kimura,
  ``Non-holomorphic multi-matrix gauge invariant operators based on Brauer
  algebra,''
  arXiv:0910.2170 [hep-th].

\bibitem{Ramgoolam:2008yr}
  S.~Ramgoolam,
  ``Schur-Weyl duality as an instrument of Gauge-String duality,''
  AIP Conf.\ Proc.\  {\bf 1031}, 255 (2008)
  [arXiv:0804.2764 [hep-th]].

\bibitem{Balasubramanian:2001nh}
  V.~Balasubramanian, M.~Berkooz, A.~Naqvi and M.~J.~Strassler,
  ``Giant gravitons in conformal field theory,''
  JHEP {\bf 0204}, 034 (2002)
  [arXiv:hep-th/0107119].

\bibitem{'tHooft:1973jz}
  G.~'t Hooft,
  ``A Planar Diagram Theory for Strong Interactions,''
  Nucl.\ Phys.\  B {\bf 72}, 461 (1974);\\
  G.~'t Hooft,
  ``A Two-Dimensional Model For Mesons,''
  Nucl.\ Phys.\  B {\bf 75}, 461 (1974).

\bibitem{Koch:2008ah}
  R.~de Mello Koch,
  ``Geometries from Young Diagrams,''
  JHEP {\bf 0811}, 061 (2008)
  [arXiv:0806.0685 [hep-th]].

\bibitem{Koch:2008cm}
  R.~de Mello Koch, N.~Ives and M.~Stephanou,
  ``Correlators in Nontrivial Backgrounds,''
  Phys.\ Rev.\  D {\bf 79}, 026004 (2009)
  [arXiv:0810.4041 [hep-th]].

\bibitem{Koch:2009jc}
  R.~de Mello Koch, T.~K.~Dey, N.~Ives and M.~Stephanou,
  ``Correlators Of Operators with a Large R-charge,''
  arXiv:0905.2273 [hep-th].

\bibitem{Lin:2004nb}
  H.~Lin, O.~Lunin and J.~M.~Maldacena,
  ``Bubbling AdS space and 1/2 BPS geometries,''
  JHEP {\bf 0410}, 025 (2004)
  [arXiv:hep-th/0409174].

\bibitem{Balasubramanian:2005mg}
  V.~Balasubramanian, J.~de Boer, V.~Jejjala and J.~Simon,
  ``The library of Babel: On the origin of gravitational thermodynamics,''
  JHEP {\bf 0512}, 006 (2005)
  [arXiv:hep-th/0508023],\\
  V.~Balasubramanian, V.~Jejjala and J.~Simon,
  ``The library of Babel,''
  Int.\ J.\ Mod.\ Phys.\  D {\bf 14}, 2181 (2005)
  [arXiv:hep-th/0505123].

\bibitem{shahin}
  V.~Balasubramanian, J.~de Boer, V.~Jejjala and J.~Simon,
  ``Entropy of near-extremal black holes in AdS$_5$,''
  JHEP {\bf 0805}, 067 (2008)
  [arXiv:0707.3601 [hep-th]],\\
  R.~Fareghbal, C.~N.~Gowdigere, A.~E.~Mosaffa and M.~M.~Sheikh-Jabbari,
  ``Nearing Extremal Intersecting Giants and New Decoupled Sectors in N = 4
  SYM,''
  JHEP {\bf 0808}, 070 (2008)
  [arXiv:0801.4457 [hep-th]].

\bibitem{Maldacena:1997re}
  J.~M.~Maldacena,
  ``The large N limit of superconformal field theories and supergravity,''
  Adv.\ Theor.\ Math.\ Phys.\  {\bf 2}, 231 (1998)
  [Int.\ J.\ Theor.\ Phys.\  {\bf 38}, 1113 (1999)]
  [arXiv:hep-th/9711200];\\
  S.~S.~Gubser, I.~R.~Klebanov and A.~M.~Polyakov,
  ``Gauge theory correlators from non-critical string theory,''
  Phys.\ Lett.\ B {\bf 428}, 105 (1998)
  [arXiv:hep-th/9802109];\\
  E.~Witten,
  ``Anti-de Sitter space and holography,''
  Adv.\ Theor.\ Math.\ Phys.\  {\bf 2}, 253 (1998)
  [arXiv:hep-th/9802150].

\bibitem{Skenderis:2007yb}
  K.~Skenderis and M.~Taylor,
  ``Anatomy of bubbling solutions,''
  JHEP {\bf 0709}, 019 (2007)
  [arXiv:0706.0216 [hep-th]].

\bibitem{Berenstein:2002jq}
  D.~E.~Berenstein, J.~M.~Maldacena and H.~S.~Nastase,
  ``Strings in flat space and pp waves from N = 4 super Yang Mills,''
  JHEP {\bf 0204}, 013 (2002)
  [arXiv:hep-th/0202021].

\bibitem{Chen:2007gh}
  H.~Y.~Chen, D.~H.~Correa and G.~A.~Silva,
  ``Geometry and topology of bubble solutions from gauge theory,''
  Phys.\ Rev.\  D {\bf 76}, 026003 (2007)
  [arXiv:hep-th/0703068].

\bibitem{us}
  R.~de Mello Koch, T.~K.~Dey, N.~Ives and M.~Stephanou,
  ``Hints of Integrability Beyond the Planar Limit,''
  JHEP {\bf 1001}, 014 (2010)
  [arXiv:0911.0967 [hep-th]].

\bibitem{hai1}
  H.~Lin, A.~Morisse and J.~P.~Shock,
  ``Strings on Bubbling Geometries,''
  JHEP {\bf 1006}, 055 (2010)
  [arXiv:1003.4190 [hep-th]].

\bibitem{hai2}
  H.~Lin,
  ``Studies on 1/4 BPS and 1/8 BPS geometries,''
  arXiv:1008.5307 [hep-th].

\bibitem{McGreevy:2000cw}
  J.~McGreevy, L.~Susskind and N.~Toumbas,
  ``Invasion of the giant gravitons from anti-de Sitter space,''
  JHEP {\bf 0006}, 008 (2000)
  [arXiv:hep-th/0003075];\\
  M.~T.~Grisaru, R.~C.~Myers and O.~Tafjord,
  ``SUSY and Goliath,''
  JHEP {\bf 0008}, 040 (2000)
  [arXiv:hep-th/0008015];\\
  A.~Hashimoto, S.~Hirano and N.~Itzhaki,
  ``Large branes in AdS and their field theory dual,''
  JHEP {\bf 0008}, 051 (2000)
  [arXiv:hep-th/0008016].

\bibitem{Berenstein:2004kk}
  D.~Berenstein,
  ``A toy model for the AdS/CFT correspondence,''
  JHEP {\bf 0407}, 018 (2004)
  [arXiv:hep-th/0403110].

\bibitem{Koch:2010gp}
  R.~d.~M.~Koch, G.~Mashile and N.~Park,
  ``Emergent Threebrane Lattices,''
  Phys.\ Rev.\  D {\bf 81}, 106009 (2010)
  [arXiv:1004.1108 [hep-th]].

\bibitem{Mikhailov:2000ya}
  A.~Mikhailov,
  ``Giant gravitons from holomorphic surfaces,''
  JHEP {\bf 0011}, 027 (2000)
  [arXiv:hep-th/0010206].

\bibitem{Beasley:2002xv}
  C.~E.~Beasley,
  ``BPS branes from baryons,''
  JHEP {\bf 0211}, 015 (2002)
  [arXiv:hep-th/0207125].

\bibitem{Kinney:2005ej}
  J.~Kinney, J.~M.~Maldacena, S.~Minwalla and S.~Raju,
  ``An index for 4 dimensional super conformal theories,''
  Commun.\ Math.\ Phys.\  {\bf 275}, 209 (2007)
  [arXiv:hep-th/0510251].

\bibitem{Biswas:2006tj}
  I.~Biswas, D.~Gaiotto, S.~Lahiri and S.~Minwalla,
  ``Supersymmetric states of N = 4 Yang-Mills from giant gravitons,''
  JHEP {\bf 0712}, 006 (2007)
  [arXiv:hep-th/0606087].

\bibitem{Mandal:2006tk}
  G.~Mandal and N.~V.~Suryanarayana,
  ``Counting 1/8-BPS dual-giants,''
  JHEP {\bf 0703}, 031 (2007)
  [arXiv:hep-th/0606088].

\bibitem{jurgis}
  J.~Pasukonis and S.~Ramgoolam,
  ``From counting to construction of BPS states in N=4 SYM,''
  arXiv:1010.1683 [hep-th].

\bibitem{Balasubramanian:2002sa}
  V.~Balasubramanian, M.~x.~Huang, T.~S.~Levi and A.~Naqvi,
  ``Open strings from N = 4 super Yang-Mills,''
  JHEP {\bf 0208}, 037 (2002)
  [arXiv:hep-th/0204196],\\
  O. Aharony, Y.E. Antebi, M. Berkooz and R. Fishman, ``Holey sheets: Pfaffians and 
  subdeterminants as D-brane operators in large $N$ gauge theories,''
  JHEP {\bf 0212}, 096 (2002)
  [arXiv:hep-th/0211152].

\bibitem{Sadri:2003mx}
  D.~Sadri and M.~M.~Sheikh-Jabbari,
  ``Giant hedge-hogs: Spikes on giant gravitons,''
  Nucl.\ Phys.\  B {\bf 687}, 161 (2004)
  [arXiv:hep-th/0312155].

\bibitem{Berenstein:2003ah}
  D.~Berenstein,
  ``Shape and holography: Studies of dual operators to giant gravitons,''
  Nucl.\ Phys.\  B {\bf 675}, 179 (2003)
  [arXiv:hep-th/0306090].

\bibitem{Berenstein:2006qk}
  D.~Berenstein, D.~H.~Correa and S.~E.~Vazquez,
  ``A study of open strings ending on giant gravitons, spin chains and
  integrability,''
  [arXiv:hep-th/0604123],\\
  D.~Berenstein and S.~E.~Vazquez,
  ``Integrable open spin chains from giant gravitons,''
  JHEP {\bf 0506}, 059 (2005)
  [arXiv:hep-th/0501078],\\
  D.~Berenstein, D.~H.~Correa and S.~E.~Vazquez,
  ``Quantizing open spin chains with variable length: An example from giant
  gravitons,''
  Phys.\ Rev.\ Lett.\  {\bf 95}, 191601 (2005)
  [arXiv:hep-th/0502172],\\
  D.~H.~Correa and G.~A.~Silva,
  ``Dilatation operator and the Super Yang-Mills duals of open strings on AdS
  Giant Gravitons,''
  [arXiv:hep-th/0608128].

\bibitem{yusuke}
  Y.~Kimura,
  ``Quarter BPS classified by Brauer algebra,''
  JHEP {\bf 1005}, 103 (2010)
  [arXiv:1002.2424 [hep-th]].

\bibitem{heslop}
  E.~D'Hoker and A.~V.~Ryzhov,
  ``Three-point functions of quarter BPS operators in N = 4 SYM,''
  JHEP {\bf 0202}, 047 (2002)
  [arXiv:hep-th/0109065],\\
  E.~D'Hoker, P.~Heslop, P.~Howe and A.~V.~Ryzhov,
  ``Systematics of quarter BPS operators in N = 4 SYM,''
  JHEP {\bf 0304}, 038 (2003)
  [arXiv:hep-th/0301104],\\
  P.~J.~Heslop and P.~S.~Howe,
  ``OPEs and 3-point correlators of protected operators in N = 4 SYM,''
  Nucl.\ Phys.\  B {\bf 626}, 265 (2002)
  [arXiv:hep-th/0107212].

\bibitem{Brown:2008rs}
  T.~W.~Brown,
  ``Permutations and the Loop,''
  arXiv:0801.2094 [hep-th].

\bibitem{tomyusuke}
T.~W.~Brown,
  ``Cut-and-join operators and N=4 super Yang-Mills,''
  arXiv:1002.2099 [hep-th],\\

\bibitem{Huang:2010ne}
  M.~x.~Huang,
  ``Higher Genus BMN Correlators: Factorization and Recursion Relations,''
  arXiv:1009.5447 [hep-th].

\bibitem{Beisert:2003tq}
  N.~Beisert, C.~Kristjansen and M.~Staudacher,
  ``The dilatation operator of N = 4 super Yang-Mills theory,''
  Nucl.\ Phys.\  B {\bf 664}, 131 (2003)
  [arXiv:hep-th/0303060].

\bibitem{de Mello Koch:2004ws}
  R.~de Mello Koch and R.~Gwyn,
  ``Giant graviton correlators from dual SU(N) super Yang-Mills theory,''
  JHEP {\bf 0411}, 081 (2004)
  [arXiv:hep-th/0410236].

\bibitem{Bhattacharyya:2008rc}
  R.~Bhattacharyya, R.~de Mello Koch and M.~Stephanou,
  ``Exact Multi-Restricted Schur Polynomial Correlators,''
  arXiv:0805.3025 [hep-th].

\bibitem{Das:2000st}
  S.~R.~Das, A.~Jevicki and S.~D.~Mathur,
  ``Vibration modes of giant gravitons,''
  Phys.\ Rev.\  D {\bf 63}, 024013 (2001)
  [arXiv:hep-th/0009019].

\bibitem{Okunkov}
Andrei Okounkov and Anatoly Vershik,
``A new approach to representation theory of symmetric groups,'' 
Selecta Mathematica, {\bf  2}, 581-605,\\
Andrei Okounkov and Anatoly Vershik,
``A New Approach to the Representation Theory of the Symmetric Groups II,''
Journal of Mathematical Sciences 
{\bf 131}, 5471-5494. 

\bibitem{Myers:1999ps}
  R.~C.~Myers,
  ``Dielectric-branes,''
  JHEP {\bf 9912}, 022 (1999)
  [arXiv:hep-th/9910053].

\bibitem{bhw}
W. Carlson, R. de Mello Koch and H. Lin, work in progress.

\end{thebibliography}
\end{document}